\documentclass{emulateapj}
\usepackage{graphicx}
\usepackage[normalem]{ulem} 
\pdfoutput=1

\def\gsim{\mathrel{\raise0.35ex\hbox{$\scriptstyle >$}\kern-0.6em\lower0.40ex\hbox{{$\scriptstyle \sim$}}}}
\def\lsim{\mathrel{\raise0.35ex\hbox{$\scriptstyle <$}\kern-0.6em\lower0.40ex\hbox{{$\scriptstyle \sim$}}}}

\begin{document}

\title{Kiloparsec--scale dust disks in high-redshift luminous submillimeter galaxies}

\shorttitle{Kpc--scale dust disks in high-z SMGs}
\shortauthors{Hodge et al.}

\author{J. A. Hodge\altaffilmark{1}}
\altaffiltext{1}{Leiden Observatory, Leiden University, P.O. Box 9513, 2300 RA Leiden, the Netherlands}
\email{hodge@strw.leidenuniv.nl}


\author{A. M. Swinbank\altaffilmark{2,3}}
\altaffiltext{2}{Centre for Extragalactic Astronomy, Department of Physics, Durham University, South Road, Durham, DH1 3LE, UK}
\altaffiltext{3}{Institute for Computational Cosmology, Durham University, South Road, Durham, DH1 3LE, UK}

\author{J. M. Simpson\altaffilmark{4}}
\altaffiltext{4}{Institute for Astronomy, University of Edinburgh, Blackford Hill, Edinburgh EH9 3HJ, UK}

\author{I. Smail\altaffilmark{2,3}}

\author{F. Walter\altaffilmark{5}}
\altaffiltext{5}{Max--Planck Institut f\"ur Astronomie, K\"onigstuhl 17, 69117 Heidelberg, Germany}

\author{D. M. Alexander\altaffilmark{2}}

\author{F. Bertoldi\altaffilmark{6}}
\altaffiltext{6}{Argelander--Institute of Astronomy, Bonn University, Auf dem H\"ugel 71, D--53121 Bonn, Germany}

\author{A. D. Biggs\altaffilmark{7}}
\altaffiltext{7}{European Southern Observatory, Karl--Schwarzschild Strasse 2, D--85748 Garching, Germany}

\author{W. N. Brandt\altaffilmark{8,9,10}}
\altaffiltext{8}{Department of Astronomy \& Astrophysics, 525 Davey Lab, The Pennsylvania State University, University Park, Pennsylvania, 16802, USA}
\altaffiltext{9}{Institute for Gravitation and the Cosmos, The Pennsylvania State University, University Park, PA 16802, USA}
\altaffiltext{10}{Department of Physics, 104 Davey Laboratory, The Pennsylvania State University, University Park, PA 16802, USA}

\author{S. C. Chapman\altaffilmark{11}}
\altaffiltext{11}{Department of Physics and Atmospheric Science, Dalhousie University, 6310 Coburg Road, Halifax, NS B3H 4R2, Canada}

\author{C. C. Chen\altaffilmark{2}}

\author{K. E. K. Coppin\altaffilmark{12}}
\altaffiltext{12}{Department of Physics, McGill University, 3600 Rue University, Montreal, QC H3A 2T8, Canada}

\author{P. Cox\altaffilmark{13}}
\altaffiltext{13}{Joint ALMA Observatory - ESO, Av. Alonso de C«ordova, 3104, Santiago, Chile}

\author{A. C. Edge\altaffilmark{2}}

\author{T. R. Greve\altaffilmark{14}}
\altaffiltext{14}{University College London, Department of Physics \& Astronomy, Gower Street, London, WC1E 6BT, UK}

\author{R. J. Ivison\altaffilmark{4,15}}
\altaffiltext{15}{UK Astronomy Technology Center, Science and Technology Facilities Council, Royal Observatory, Blackford Hill, Edinburgh EH9 3HJ, UK}

\author{A. Karim\altaffilmark{6}}

\author{K. K. Knudsen\altaffilmark{16}}
\altaffiltext{16}{Department of Earth and Space Sciences, Chalmers University of Technology, Onsala Space Observatory, SE--43992 Onsala, Sweden}

\author{K. M. Menten\altaffilmark{6}}

\author{H.--W. Rix\altaffilmark{5}}

\author{E. Schinnerer\altaffilmark{5}}

\author{J. L. Wardlow\altaffilmark{2}}

\author{A. Weiss\altaffilmark{17}}
\altaffiltext{17}{Max--Planck Institut f\"ur Radioastronomie, Auf dem H\"ugel 69, D--53121 Bonn, Germany}

\author{P. van der Werf\altaffilmark{1}}

\begin{abstract}
\noindent We present high--resolution (0.16$''$) 870$\mu$m Atacama Large Millimeter/submillimeter Array (ALMA) imaging of 16 luminous ($L_{\rm IR} \sim 4 \times 10^{12} L_{\odot}$) submillimeter galaxies (SMGs) from the ALESS survey of the Extended \textit{Chandra} Deep Field South.  This dust imaging traces the dust--obscured star formation in these $z\sim2.5$ galaxies on $\sim$1.3\,kpc scales. The emission has a median effective radius of $R_e=0.24$$''$$\pm$0.02$''$, corresponding to a typical physical size of $R_{e}=1.8\pm$0.2\,kpc. We derive a median S\'ersic index of $n=0.9$$\pm$0.2, implying that the dust emission is remarkably disk-like at the current resolution and sensitivity. We use different weighting schemes with the visibilities to search for clumps on 0.12$''$ ($\sim$1.0\,kpc) scales, but we find no significant evidence for clumping in the majority of cases. Indeed, we demonstrate using simulations that the observed morphologies are generally consistent with smooth exponential disks, suggesting that caution should be exercised when identifying candidate clumps in even moderate S/N interferometric data. We compare our maps to comparable--resolution \textit{HST} \textit{H}$_{\rm 160}$-band images, finding that the stellar morphologies appear significantly more extended and disturbed, and suggesting that major mergers may be responsible for driving the formation of the compact dust disks we observe. The stark contrast between the obscured and unobscured morphologies may also have implications for SED fitting routines that assume the dust is co-located with the optical/near--IR continuum emission. Finally, we discuss the potential of the current bursts of star formation to transform the observed galaxy sizes and light profiles, showing that the $z\sim0$ descendants of these SMGs are expected to have stellar masses, effective radii, and gas surface densities consistent with the most compact massive (M$_{*}\sim$ 1--2$\times$10$^{11}$ M$_{\odot}$) early--type galaxies observed locally.


\end{abstract}

\section{INTRODUCTION}
\label{Intro}

How high--redshift galaxies formed their stars remains an open question. 
Deep (rest--frame) UV/optical surveys have yielded 
 large samples of high--redshift ($z$$\sim$1.5--3.5) star--forming galaxies selected based on magnitude/color properties \citep[BM/BX, $BzK$; e.g.,][]{2004ApJ...604..534S, 2004ApJ...617..746D, 2007ApJ...670..173D, 2007ApJ...670..156D}, 
 the study of which has provided a basic picture of their formation. 
In particular, studies of the ionized gas kinematics in such galaxies have uncovered a high fraction of large rotating disks among the massive, optically--bright systems \citep[e.g.,][]{2006ApJ...645.1062F, 2008ApJ...682..231S, 2012MNRAS.426..935S}. These studies suggest that secular processes within star--forming galaxies are driving their gas and stars into the central regions, building up exponential disks and massive bulges without the need for major mergers \citep[e.g.,][]{2008ApJ...688...67E, 2008ApJ...687...59G, 2009ApJ...703..785D, 2013MNRAS.435..999D, 2016ASSL..418..355B}.

The most luminous galaxies at high--redshift are the dusty star--forming galaxies originally detected in the submillimeter and known as submillimeter galaxies \citep[SMG; e.g.,][]{2002PhR...369..111B, 2005ARA&A..43..677S, 2013ARA&A..51..105C, 2014arXiv1402.1456C}.
 Their large luminosities ($L_{\rm IR} > 10^{12-13}$ L$_{\odot}$, qualifying them as ultra-- or even hyper--luminous infrared galaxies) 
make them easier to observe in the distant universe, in principle, though whether their star formation process differs from less extreme galaxies is still debated. The canonical picture is that the majority of SMGs are scaled--up ultra-luminous infrared galaxies \citep[ULIRGs;][]{1996ARA&A..34..749S} -- i.e., starburst--dominated major mergers  \citep[e.g.,][]{2010MNRAS.401.1613N}, 
although non-cosmological hydrodynamic simulations have suggested that SMGs could be a heterogeneous population: a mix of pre-merger pairs of disk galaxies, merger--induced starbursts, and isolated gas--rich disk galaxies undergoing a secular burst \citep[e.g.,][]{2011ApJ...743..159H, 2012MNRAS.424..951H}. 
Still other models posit that the submillimeter-luminous phase is long-lived and associated with the bombardment of a central halo by numerous sub-halos in early Universe proto-clusters \citep{2015Natur.525..496N}.
Finally, some models propose that SMGs may simply represent the most massive extension of the normal $z>2$ star--forming galaxy population \citep[e.g.,][]{2005MNRAS.363....2K, 2009MNRAS.395..160K, 2009MNRAS.396.2332K, 2010MNRAS.404.1355D}. This last theory may be at odds with claims that normal (BM/BX, $BzK$) high--redshift star--forming galaxies seem to follow a different sequence than SMGs on the $M_{\rm gas}$/$L_{\rm IR}$ plane 
(\citealt[e.g.,][]{2010MNRAS.407.2091G, 2010ApJ...713..686D, 2015ApJ...798L..18H}; although see \citealt{2011MNRAS.412.1913I}).

In order to better understand how SMGs fit into the larger evolutionary picture
-- and, more broadly, how star formation occurred in high--redshift galaxies in general -- 
resolved observations of the spatial distribution of the star formation are essential. 
However, studies based solely
in the (rest--frame) optical/UV \citep[e.g.,][]{2003ApJ...599...92C, 2005ApJ...622..772C, 2010MNRAS.405..234S, 2015ApJ...799..194C} must contend with dust--obscuration, which can make such emission challenging to detect in the most highly star--forming galaxies, 
and where patchy reddening could potentially affect the apparent morphology, particularly in the rest--frame UV. 
Some studies therefore use the Plateau de Bure Interferometer (PdBI) and Karl G. Jansky Very Large Array (VLA) to target radio synchrotron emission, a potential proxy for star formation, 
 or molecular line emission (CO), which traces the gas reservoirs required to fuel star formation, at sub-arcsecond resolution \citep[$\gtrsim$0.2$^{\prime\prime}$; e.g.,][]{2010Natur.463..781T, 2010ApJ...724..233E, 2010MNRAS.405..219B, 2012ApJ...760...11H, 2013ApJ...776...22H, 2013ApJ...768...74T, 2013ApJ...773...68G, 2014MNRAS.442..558A, 2015ApJ...809..175B, 2015A&A...584A..32M}. The molecular gas studies in particular reveal large clumpy disks in both the more `normal' high--redshift galaxies and even in some SMGs \citep{2012ApJ...760...11H}, in apparent agreement with claims of $\sim$kpc-scale star-forming regions in high-redshift galaxies from the rest-frame optical/UV \citep[e.g.,][]{2004ApJ...603...74E, 2011ApJ...739...45F, 2012ApJ...757..120G, 2015ApJ...800...39G} and H$\alpha$ line emission \citep{2011ApJ...733..101G}. 
Such massive kpc--scale clumps are thought to form in--situ by gravitational instability due to the gas--richness of these high--redshift galaxies \citep[e.g.,][]{2009ApJ...703..785D, 2014ApJ...780...57B}. 
Moreover, molecular gas observations can also provide valuable information on the kinematics of the systems. 
For example, based on observations of continuum and various CO transitions (up to CO[7--6]) in a sample of 12 SMGs, \citet{2010ApJ...724..233E} suggested that practically all SMGs are major mergers.
However, such studies have been very expensive observationally, and in many cases at best marginally resolve the sources (see \citealt{2013ARA&A..51..105C} for a review). 

A more direct way to trace the obscured star--forming regions in high--redshift galaxies is through observations of the dust continuum emission in the rest--frame far-infrared (FIR), corresponding to observed submillimeter wavelengths for sources at $z>1$. The FIR dust continuum is dominantly powered by recently--formed, massive stars, making it an excellent tracer of the bolometric luminosity -- and thus star formation -- in dusty starbursts such as SMGs.
While the resolution achievable by early submillimeter interferometric observations \citep[e.g.,][]{2008ApJ...673L.127D, 2011ApJ...726L..18W, 2012A&A...548A...4S, 2013ApJ...768...91H} was too poor ($>$1$^{\prime\prime}$) to sufficiently resolve high--redshift galaxies except for in a handful of cases \citep[e.g.,][]{2008ApJ...688...59Y, 2012ApJ...760...11H}, 
recently there have been some first attempts to constrain the sizes of larger samples of SMGs -- as well as massive dusty star--forming galaxies selected as likely progenitors of $z\sim2$ compact quiescent galaxies
 -- in the submillimeter \citep[e.g.,][]{2015ApJ...799...81S, 2015ApJ...810..133I, 2016ApJ...827L..32B}, revealing compact ($R_{e}\sim1$\,kpc) dusty starbursts. However, how this star formation is distributed within the sources --
e.g., whether it lies in clumpy disks or is strongly centrally peaked due to the violent and dissipative collapse expected from major merger remnants \citep{2011ApJ...730....4B} -- is still unknown. 
Moreover, only in rare cases of gravitational magnification \citep{2010Natur.464..733S, 2015PASJ...67...93H}
or case studies of single extreme sources \citep{2015ApJ...798L..18H, 2016ApJ...827...34O} have individual star--forming regions in an SMG -- or any high--redshift galaxy -- been potentially resolved in the FIR. 
While seemingly consistent with the kpc--scale clumps observed in the rest--frame optical/UV and H$\alpha$/CO line emission, 
the reality of these low--S/N ``clumps'' -- which are argued to play a key role in high--redshift galaxy formation and evolution -- 
has not yet been confirmed.

With ALMA, the situation is now fundamentally changed.
The long baselines and large number of antennas make it possible to resolve the star-forming regions in galaxies on scales of $\lesssim$1 kpc, similar to the resolution achievable for nearby galaxies with \emph{Herschel}, and at a sensitivity sufficient to map the morphology of the emission.  
We therefore used ALMA to conduct high--resolution ($\sim$0.16$''$ FWHM)
 Band 7 (344\,GHz)
  mapping of the (rest--frame) FIR--continuum in 17 SMGs selected from our ALMA Cycle~0 compact configuration survey of single--dish 344\,GHz LABOCA sources detected in the
   Extended \textit{Chandra} Deep Field South (ECDFS) by \citet{2009ApJ...707.1201W}, constituting the largest, most homogenous, and highest--sensitivity sample of interferometrically observed SMGs to date \citep[ALESS;][]{2013ApJ...768...91H, 2013MNRAS.432....2K}.

We begin in \S\ref{data} with the details of the observations.  Our results are presented in \S\ref{results}, followed by a discussion in \S\ref{discussion}. We summarize our conclusions in \S\ref{summary}. Where applicable we assume a concordance, flat $\Lambda$CDM cosmology of H$_0$=71 km\,s$^{-1}$ Mpc$^{-1}$, $\Omega_{\Lambda}$=0.73, and $\Omega_{M}$=0.27 \citep{2003ApJS..148..175S, 2007ApJS..170..377S}.  All magnitudes are on the AB system.

\section{Observations and Data Reduction}
\label{data}

\subsection{Sample Selection \& Observations}
\label{obs}
\begin{figure}
\centering
\includegraphics[scale=0.43,trim={0 0 0 0cm},clip]{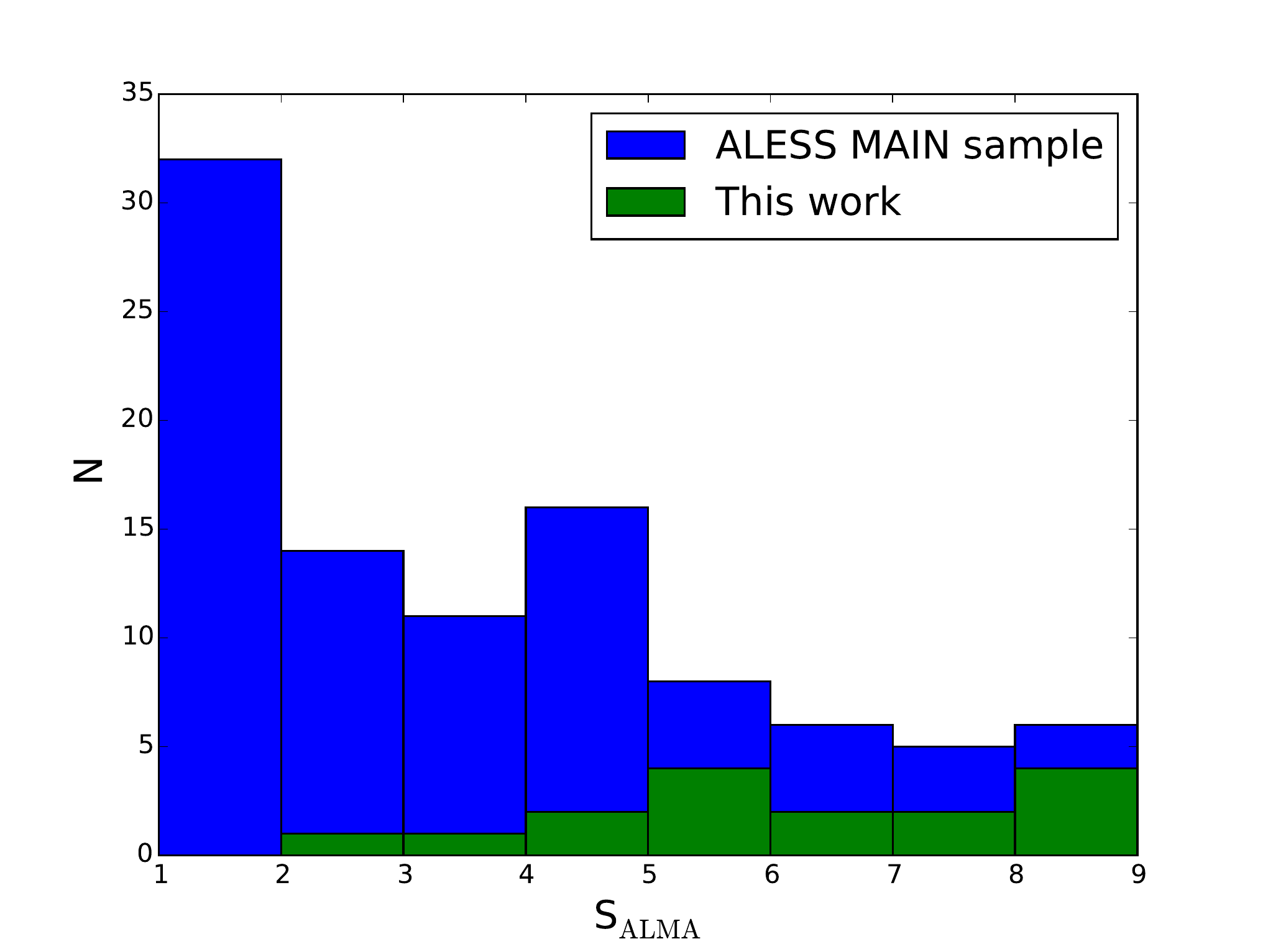}
\caption{ALMA 870$\mu$m flux density for the high--resolution sources targeted in this paper compared to the entire ALESS
 MAIN sample. The flux densities have been corrected for the effect of flux boosting -- see \S\ref{flux}. The high--resolution targets of this study were chosen from the randomly--selected \textit{HST}--covered fields and preferentially target the 
 brighter ALMA SMGs in this field. 
  }
\label{fig:fluxhist}
\end{figure}

The ALMA observations analyzed here were taken between 11--27 Aug 2015 as part of our rolled--over Cycle~1 Project \#2012.1.00307.S.
We targeted 15 fields from our Cycle~0 ALESS survey \citep{2013ApJ...768...91H}, which itself observed 122 of the 126 single--dish--selected submillimeter sources originally detected in the LESS survey of the ECDFS \citep{2009ApJ...707.1201W}. The 15 fields were selected from $\sim$40 fields which, as of the Cycle~1 proposal deadline in early 2012, had either existing or forthcoming 
deep \textit{Hubble Space Telescope (HST)} observations through CANDELS or our 
Cycle 20 \textit{HST} program \citep{2015ApJ...799..194C}. 
 Specifically, we selected the fields containing the submillimeter--brightest ALMA SMGs from the \textit{HST}-covered fields, which were themselves randomly selected. Although some of the ALESS SMGs may be marginally resolved in the $\sim$1.6$''$
  (FWHM) Cycle~0 data along one or more axes \citep[and only one source definitively so;][]{2013ApJ...768...91H}, no selection was made on source extent or morphology in the ALMA or \textit{HST} images so as to avoid biasing the results. 
Four of the SMGs are associated with X--ray sources \citep[ALESS 17.1, 45.1, 67.1, and 73.1;][]{2013ApJ...778..179W}. 
The flux density distribution for the sources targeted in this program compared with that for the entire ALESS Cycle~0 sample is shown in Figure~\ref{fig:fluxhist}, where we see that the sources targeted in this study are slightly brighter than the average SMGs. 

As in our Cycle~0 ALESS program, we observed all fields with ALMA's Band 7 centered at 344 GHz/870$\mu$m to facilitate direct comparison of the measured flux densities.
We utilized the ``single continuum'' spectral mode, 
with 4 $\times$ 128 dual polarization channels over the 8 GHz bandwidth.
At this frequency, ALMA has a 17.3$^{\prime\prime}$ primary beam (FWHM).

Three fields (LESS 1, 15 and 67) contained multiple SMGs detected in the Cycle~0 {\sc main} ALESS catalog at 1.6$''$ resolution, and four fields contained SMGs from the Cycle~0 {\sc supplementary} catalog in addition to the primary source(s) from the {\sc main} catalog  \citep{2013ApJ...768...91H}. 
In all cases except for LESS 1, the ALMA beam was centered on the brightest Cycle~0 ALESS source in the field in order to maximize sensitivity for the high-resolution observations.
As a result, the majority of the Cycle~0 {\sc supplementary} sources fall outside the coverage of the ALMA beam. The observations presented here thus include 18 SMGs from the Cycle~0 {\sc main} catalog and one SMG from the Cycle~0 {\sc supplementary} catalog, or 19 SMGs in total (within the 17.3$''$ FWHM of the primary beam). 

The ALMA observations were requested in the C32-6 configuration and carried out with 46 antennas in an extended configuration 
(minimum baseline of $\sim$15m,
maximum baseline of $\sim$1.6 km). The phase, flux, and bandpass calibrators were J0348$-$2749, J0334$-$401, and J0522$-$3627, respectively, and the total integration time on each of the target fields was approximately 8 minutes.
The phase stability/weather conditions were good, with a median PWV at zenith of $\sim$0.7 mm. 

\begin{figure*}
\centering
\includegraphics[scale=0.73,trim={0 0 0 1cm},clip]{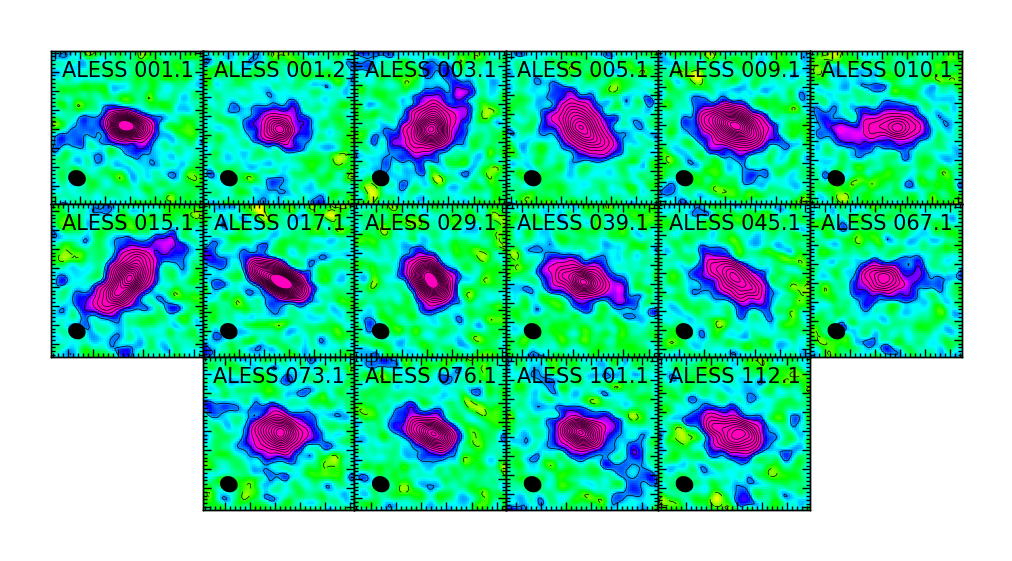}
\vspace{-10mm}
\caption{ALMA images (each 1.6$''$$\times$1.6$''$, or 13 kpc at $z\sim2.5$) of the 870$\mu$m emission from 16 SMGs at 0.17$''$$\times$0.15$''$ resolution, corresponding to a physical scale of 1.4$\times$1.2 kpc at $z\sim2.5$ (beamsize shown in the bottom left corners). Contours go from $\pm$2--30$\sigma$ in steps of 2$\sigma$, and the typical RMS ($\sigma\sim64$ $\mu$Jy beam$^{-1}$) corresponds to a rest--frame brightness temperature of T$_{B} = 0.09$ K at $z\sim2.5$. Major tick marks indicate 0.2$''$. The extended
dust emission in these galaxies is 
distributed over a $\sim$few kpc scales and
smooth and disk-like at 
our sensitivity and resolution.
We note that the source positions (and/or stellar environments) of ALESS 5.1 and 10.1  suggest these sources are potentially weakly lensed (see also Figure~\ref{fig:thumbs} and \S\ref{HSTcomp}).}
\label{fig:hires}
\end{figure*}

\subsection{Data Reduction \& Imaging }
\label{reduction}
The ALMA data were reduced using the Common Astronomy Software Application\footnote{http://casa.nrao.edu} ({\sc casa}) version 4.3.1.
The delivered reduction produced $uv$--data products of high quality and was therefore used without further modifications. 
The $uv$--data were imaged using {\sc casa} version 4.3.1, with subsequent analysis carried out in {\sc casa} version 4.5.0. 

Imaging was carried out using the {\sc clean} algorithm with a variety of different weightings and $uv$--taperings to explore the extent to which the sources were resolved by the observations and the total flux densities were recovered (see \S\ref{flux}). 
The (compact configuration) Cycle~0 data were not co--added to the new data given the much poorer data quality and (in some cases) offset pointing centers.
For the untapered maps, multi-scale {\sc clean} \citep{2008ISTSP...2..793C} was employed using scales of [0$''$, 0.3$''$, 0.6$''$, 1.2$''$]. While the largest scale was set to approximately encompass the largest coherent structure visible in the maps, 
we found that the specific number and distribution of these scales did not significantly affect the results, in agreement with other studies \citep[e.g.,][]{2008AJ....136.2897R}.

All maps were cleaned interactively using 1.5$''$ circular regions around sources with emission in clear excess ($\sim$4--5$\sigma$) of the residuals. These sources were cleaned down to $\sim$2.5$\sigma$, a process which typically required 1--5 major clean cycles of 50 iterations each.  
The resulting images are 25.6$^{\prime\prime}$ per side and have a pixel scale of 0.02$^{\prime\prime}$,
and the naturally weighted maps achieve a typical synthesized beam of 0.17$''$$\times$0.15$''$ and RMS noise of $\sim$64 $\mu$Jy beam$^{-1}$, corresponding to a rest--frame brightness temperature of T$_{B} = 0.09$ K at $z\sim2.5$. A set of maps was also produced using Briggs weighting with a robust parameter of $R=-0.5$, resulting in a resolution of 0.12$''$$\times$0.11$''$ and typical RMS noise values of $\sim$130\,$\mu$Jy beam$^{-1}$. 
We did not attempt to self-calibrate the data. 
The absolute flux calibration has an uncertainty of $\sim$10\%, and this uncertainty is not included in the error bars for individual source flux densities.

Of the 19 Cycle~0 SMGs targeted by this project, 16 were detected in the new ALMA data at very high (S/N$_{peak}$$>$10$\sigma$) significance, allowing us to investigate the distribution of their dusty star formation. 
These SMGs have flux densities ranging from S$_{870\mu m}=3.4-9.0$\,mJy in our Cycle~0 data ($\sim$1.6$''$ FWHM). 
Of the three remaining SMGs, one (ALESS 1.3) was detected at lower significance (S/N$_{peak}$$<$10$\sigma$) and two others (ALESS 15.3 and 67.2) were undetected. 
These sources had flux densities of 2.0\,mJy and 1.7\,mJy (corresponding to signal--to--noise (S/N) values of 3.8 and 4.2) in the Cycle~0 catalog, respectively, and based on the multi-wavelength data presented in \citet{2014ApJ...788..125S}, it is possible that ALESS 15.3 was spurious and ALESS 67.2 has been resolved out (see Chen et al.\,2016, in preparation). 
In Figure~\ref{fig:hires}, we show image cutouts for each of the 16 strongly detected SMGs in the naturally weighted maps (0.17$''$$\times$0.15$''$ FWHM resolution), where the extended nature of the SMGs is readily apparent. These sources span a redshift range $z=0.76-4.95$, with a median redshift ($z=2.6\pm0.5$) and infrared luminosity ($L_{\rm IR} = 3.6 \pm 0.9 \times 10^{12} L_{\odot}$) consistent with the sample as a whole \citep[$z=2.5\pm0.2$ and $L_{\rm IR} = 3.0 \pm 0.3 \times 10^{12} L_{\odot}$;][]{2014ApJ...788..125S, 2014MNRAS.438.1267S}.

\begin{figure}
\includegraphics[scale=0.46]{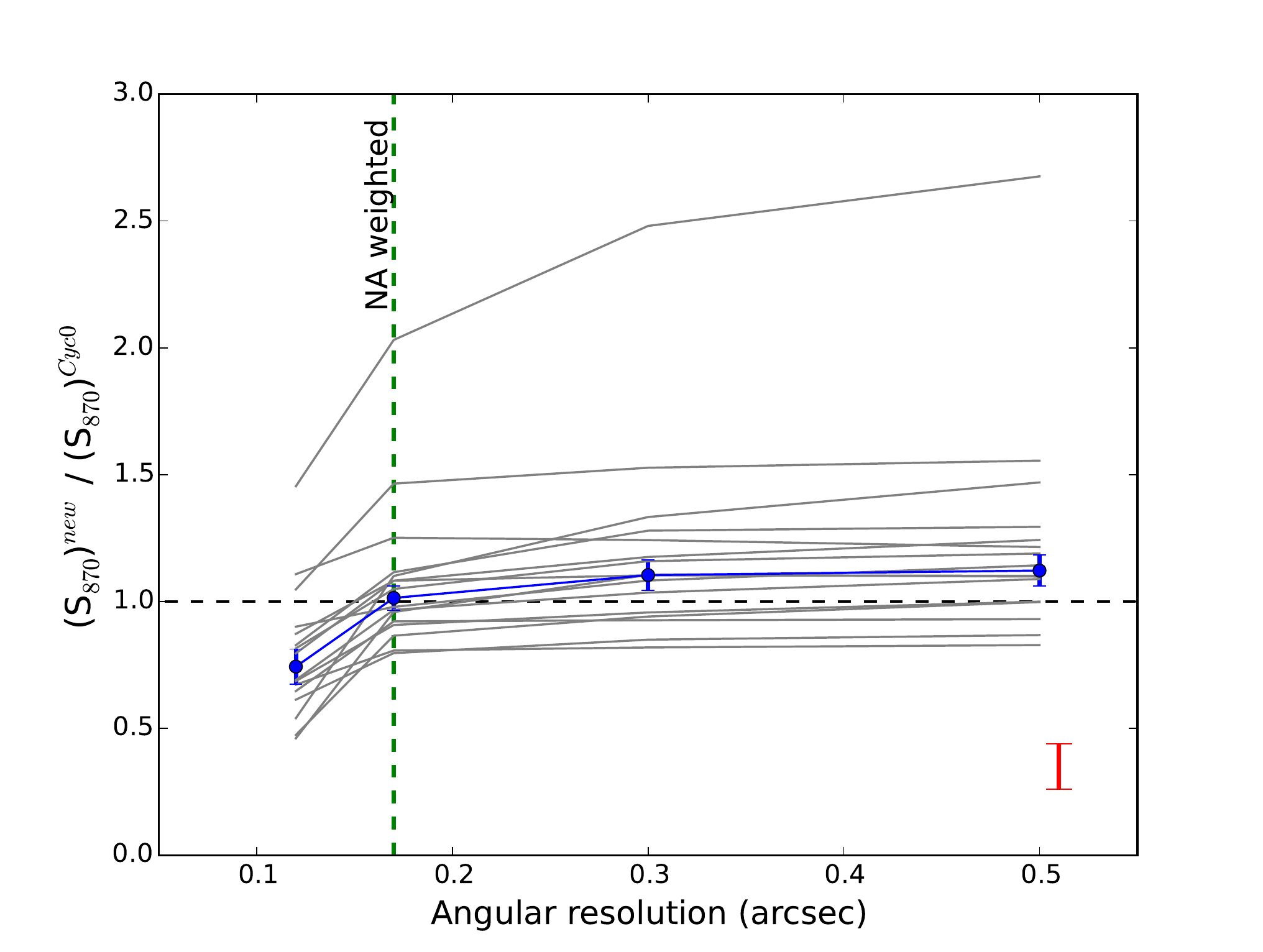}
\vspace{-2mm}
\caption{The fraction of the Cycle~0 flux density recovered for sources imaged at various spatial resolutions in our new study. The horizontal dashed line indicates a recovery fraction of 100\% compared to the earlier, low--resolution Cycle~0 data, and the vertical dashed line indicates the resolution of the naturally weighted maps. The median recovered fraction for the sample is shown by the solid blue line, and the red error bar shows the absolute flux calibration uncertainty. The highest outlier corresponds to a source (ALESS 101.1) from a lower quality ({\sc supplementary}) Cycle 0 map. While the naturally weighted images may be missing a fraction ($\sim$10--15\%) of the emission from what is presumably a more extended component ($\ge$1$''$ or $\ge$10\,kpc),
 the new ALMA observations are not formally resolving out emission, consistent with the maximum recoverable scale expected for this configuration ($\gtrsim$2$''$).}
\vspace{2mm}
\label{fig:fluxcomp}
\end{figure}

\subsection{Recovered Flux Density}
\label{flux}

In order to test whether our new, higher-resolution ALMA images recover all of the flux density from the sources, we compared the images made at various spatial resolutions with the results obtained in Cycle 0 using a more compact configuration. The Cycle~0 flux densities were taken from \citet{2013ApJ...768...91H} and have been corrected for the effect of flux boosting \cite[e.g.,][]{2015ApJ...807..128S}, which is a statistical enhancement, on average, of the measured fluxes for populations where fainter sources far outnumber the brighter ones. 
In such cases, every measurement is more likely to result from one of many fainter sources than from one of few brighter ones relative to the measurement, and the effect is most pronounced for low S/N detections.
For the new data imaged at a particular resolution, we calculated the flux density recovered by masking the emission below 2$\sigma$. For the untapered data, we then used the masks from the next lowest resolution to mask the higher-resolution images further (e.g., 0.3$''$ masks for the 0.17$''$ images; 0.17$''$ masks for the 0.12$''$ images).  This combination of steps allowed us to isolate $>$2$\sigma$ contiguous emission associated with each detected source in an automated way, which we then summed using an aperture of radius 3$\times$$b_{maj}$, where $b_{maj}$ is the FWHM (major axis) of the synthesized beam at that resolution.

Figure~\ref{fig:fluxcomp} shows the flux density recovered as a function of angular resolution  (expressed as a fraction of the Cycle~0 flux density) for individual sources and the sample median. For most sources, the recovered fraction rises steeply from the highest resolution maps ($\sim$0.1$''$; median fraction of $f=74\pm7\%$) to the naturally weighted maps ($\sim$0.16$''$; median fraction of $f=101\pm6\%$). This indicates, at face value, that the naturally weighted maps are recovering all of the flux detected in the Cycle~0 maps. However, there appears to be a potentially small increase in the recovered fraction in the $uv$--tapered data, with median fractions of $f=110\pm6\%$ and $f=112\pm6\%$ in the 0.3$''$ and 0.5$''$ maps, respectively. This modest excess may in part be due to the uncertainty in the overall flux calibration between the datasets, which, when taken into account, yields a median recovered fraction in the $uv$--tapered data consistent with the Cycle~0 values. 
As the quality of the Cycle~0 data was much poorer \citep[for example, the highest outlier in Figure~\ref{fig:fluxcomp} corresponds to {\sc supplementary} source ALESS 101.1 from a lower quality map; ][]{2013ApJ...768...91H}, we conclude that the true flux densities are better determined by the (new) tapered images. This suggests that the naturally weighted images are at most missing a fraction ($\sim$10--15\%) of the emission from what is presumably a more extended component ($\gtrsim$2$''$). We conclude that, in general, the Cycle~1 observations do not appear to be formally resolving out emission due to the array configuration, consistent with the maximum recoverable scale expected for this configuration ($\gtrsim$2$''$). We will investigate whether the emission potentially ``missing'' from the naturally weighted maps has any implications for the implied galaxy sizes in \S\ref{uvfits}.

\begin{deluxetable*}{ l c c c c c c c }
\tabletypesize{\small}
\tablewidth{14cm}
\tablecaption{ALESS SMG Observed 870$\mu$m Dust Properties \label{tab-1}}
\tablehead{
\colhead{Source ID} & \multicolumn{2}{c}{FWHM$_{\rm maj}$\tablenotemark{a}} & \colhead{$b/a$\tablenotemark{b}} & \colhead{PA\tablenotemark{c}} 
 & \colhead{FWHM$_{\rm circ}$\tablenotemark{d}} & \colhead{$R_e$\tablenotemark{e}} & \colhead{$n$\tablenotemark{f}}  \\
\colhead{ }  & \colhead{[$''$]} & \colhead{[kpc]} & \colhead{--} & \colhead{[deg]} & \colhead{[$''$]} & \colhead{--} & \colhead{--} } 
\startdata
ALESS 1.1 &  0.27$\pm$0.01  & 1.8$\pm$0.1 & 0.5 & 80$\pm$3 & 0.23$\pm$0.02 & 0.16 & 1.7 \\
ALESS 1.2 & 0.31$\pm$0.03  & 2.1$\pm$0.2 & 0.7 & 80$\pm$17 & 0.34$\pm$0.03 & 0.23  &  2.4 \\
ALESS 3.1 & 0.38$\pm$0.02 & 2.6$\pm$0.1 & 0.7 & 138$\pm$10 & 0.33$\pm$0.02 & 0.24 & 1.4 \\
ALESS 5.1 & 0.50$\pm$0.03 & 4.0$\pm$0.2 & 0.6 & 44$\pm$3 & 0.38$\pm$0.01 & 0.26 & 0.7 \\
ALESS 9.1 & 0.44$\pm$0.02 & 3.0$\pm$0.1 & 0.6 & 72$\pm$3 & 0.35$\pm$0.01 & 0.23 & 0.7 \\
ALESS 10.1 & 0.70$\pm$0.06 & 5.2$\pm$0.4 & 0.3 & 94$\pm$3 & 0.40$\pm$0.02 & 0.39 & 1.0\\
ALESS 15.1 & 0.56$\pm$0.02 & 4.8$\pm$0.2 & 0.4 & 140$\pm$2 & 0.39$\pm$0.02 & 0.31 & 0.9 \\
ALESS 17.1 & 0.40$\pm$0.02 & 3.5$\pm$0.1 & 0.3 & 62$\pm$1 & 0.29$\pm$0.02 & 0.20 & 0.5 \\
ALESS 29.1 & 0.31$\pm$0.01 & 2.7$\pm$0.1 & 0.6 & 35$\pm$3 & 0.26$\pm$0.01 & 0.16 & 0.7 \\
ALESS 39.1 & 0.47$\pm$0.03 & 3.9$\pm$0.2 & 0.4 & 73$\pm$4 & 0.32$\pm$0.02 & 0.27 & 1.2 \\
ALESS 45.1 & 0.51$\pm$0.04 & 4.3$\pm$0.3 & 0.4 & 56$\pm$2 & 0.37$\pm$0.01 & 0.26 & 0.5 \\
ALESS 67.1 & 0.44$\pm$0.04 & 3.7$\pm$0.3 & 0.5 & 89$\pm$6 & 0.32$\pm$0.02 & 0.25 & 1.2 \\
ALESS 73.1 & 0.36$\pm$0.02 & 2.4$\pm$0.1 & 0.7 & 89$\pm$9 & 0.34$\pm$0.02 & 0.20 & 1.0 \\
ALESS 76.1 & 0.33$\pm$0.02 & 2.5$\pm$0.2 & 0.5 & 64$\pm$3 & 0.27$\pm$0.01 & 0.17 & 0.6 \\
ALESS 101.1 & 0.32$\pm$0.02 & 2.5$\pm$0.2 & 0.6 & 80$\pm$7 & 0.27$\pm$0.02 & 0.27 & 2.5 \\
ALESS 112.1 & 0.45$\pm$0.03 & 3.8$\pm$0.2 & 0.6 & 70$\pm$4 & 0.36$\pm$0.01 & 0.22 & 0.5 
\enddata
\label{tab:1}
\tablenotetext{a}{FWHM of the major axis derived from a two--dimensional Gaussian fit in the image plane.}
\tablenotetext{b}{Axis ratio from the two--dimensional Gaussian fit.}
\tablenotetext{c}{Position angle from the two--dimensional Gaussian fit.}
\tablenotetext{d}{FWHM of a one--dimensional Gaussian fit to the azimuthally averaged profile in the image plane.}
\tablenotetext{e}{Effective (half--light) radius of the major axis from a two--dimensional S\'ersic profile fit. The typical error ranges from 15--27\%. }
\tablenotetext{f}{S\'ersic index from the two--dimensional S\'ersic profile fit. The typical error is in the range 26--33\%. }
\end{deluxetable*}

\begin{figure*}
\includegraphics[scale=0.62]{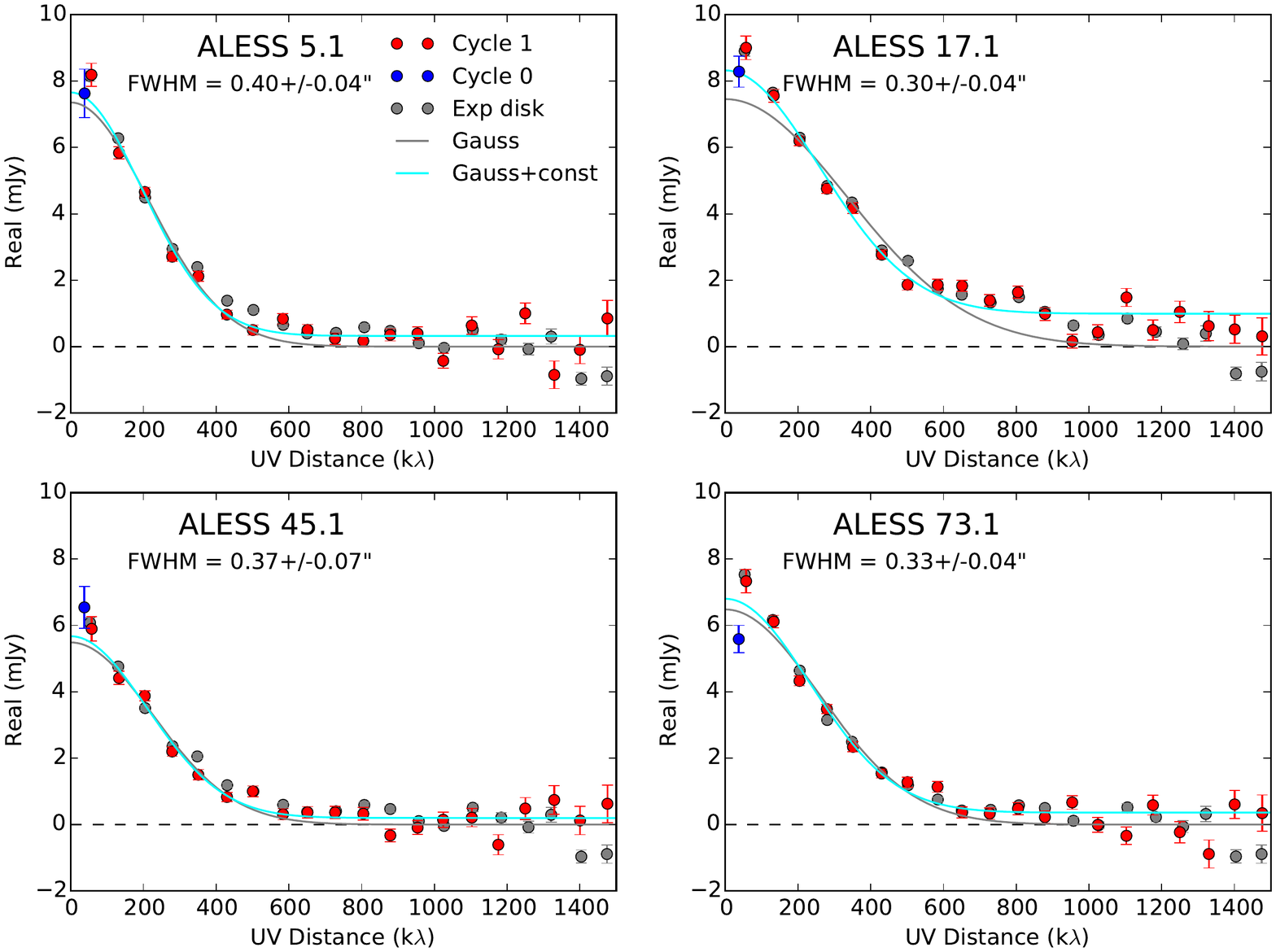}
\vspace{-6mm}
\caption{Visibility ($uv$)-profiles for four isolated SMGs from our study. The new observations (Cycle 1) have been phase-shifted to center on the source of interest and subsequently radially averaged in bins of 75 k$\lambda$ (red circles). The low-resolution Cycle~0 data had a similar procedure applied, 
and the results were then scaled by the response of the Cycle~0 primary beam at the source position (blue circles). Also shown are simulated profiles (with noise added) of exponential disks ($n=1$) with the same approximate flux densities, effective radii, axial ratios, and $uv$--coverage as the sources (gray circles). Two fits to the Cycle~1 data are shown: 1) A single Gaussian fit; and 2) a Gaussian plus constant, where the latter corresponds to a point source (or point sources) in the image plane. The FWHM listed is from the second fit. We find no evidence that our new Cycle~1 data are resolving out extended emission, in agreement with \S\ref{flux}, and we conclude that the sizes measured in the image plane are robust.
}
\label{fig:uvfits}
\end{figure*}

\section{RESULTS}
\label{results}

\subsection{The dust profiles of submillimeter galaxies}
\label{sizes}

\subsubsection{Image plane}
Figure~\ref{fig:hires} demonstrates that the dust-obscured star formation in these SMGs is extended on scales larger than our beam size (0.17$''$$\times$0.15$''$). Following \citet{2015ApJ...799...81S}, we quantified the morphology and extent of the emission by fitting each source in the image plane with three models: (1) a point source (assuming the \textsc{clean} beam); (2) a two--dimensional Gaussian; and (3) a two--dimensional S\'ersic profile. The residuals from the various fits are shown 
in Figure~\ref{fig:modelfits} in the Appendix. The point source fit is ruled out in all cases by $>$5$\sigma$ residuals. 
The parameters for the (deconvolved) two--dimensional Gaussian and S\'ersic profile fits for all SMGs are listed in Table~\ref{tab:1}. 
While many of the SMGs appear elliptical, this is most likely due to inclination and optical depth effects. As such, we report the parameters for the fits along the major axis of each source, though we also quote the axis ratios from the Gaussian fits for completeness. 

The median major axis size of the Gaussian fits is FWHM$=$0.42$''$$\pm$0.04$''$, and the median axis ratio is $b/a = 0.53\pm0.03$, where the errors on the median values were calculated via bootstrapping. The corresponding median physical size is FWHM$=$3.2$\pm$0.4\,kpc.
In the majority (9/16) of the sources, there is no significant evidence (i.e., $>$3$\sigma$ residuals from the Gaussian model) that the extra degree of freedom required for the S\'ersic profile fits is justified. The remaining sources show 3--5$\sigma$ residuals from the Gaussian model, indicating that the S\'ersic profile is preferred. The median S\'ersic profile has an index of $n=0.9$$\pm$0.2 and an effective radius of $R_e=0.24$$''$$\pm$0.02$''$, corresponding to a typical physical size of $R_{e}=1.8\pm$0.2\,kpc. 
Noting that a Gaussian fit is equivalent to a S\'ersic profile fit with $n=0.5$ and FWHM$=$2.02$\times$$R_e$, the median S\'ersic profile appears more centrally peaked than a Gaussian profile, and is consistent with an exponential disk. 
Only two SMGs (ALESS 1.2 and 101.1) have estimated S\'ersic indices $n>$2, indicating more centrally peaked emission. 
The four SMGs associated with X--ray sources (ALESS 17.1, 45.1, 67.1, and 73.1) have median parameters ($R_e=0.23$$''$$\pm$0.02$''$, $n=0.8$$\pm$0.2) consistent with the 
full
 sample. 

In order to test the robustness of the derived parameters, we inserted 10,000 model sources with S/N ratios similar to our observations into the naturally weighted maps to see how well we could recover their S\'ersic parameters. The input parameters were drawn from uniform distributions with ranges $n=0.2-5.0$, $R_e=0.1''-0.3''$, and axis ratio $b/a=0.1-1.0$. We find that the input parameters are well-recovered, with systematic biases at the $\sim$1\% level. The 1$\sigma$ scatter is a function of the input parameters, ranging from 15--27\% for the effective radius and 26--33\% for the S\'ersic index.

Finally, we create a deep composite image by combining 2$''$ cutouts centered on the source centroids. Prior to the stacking, the individual sources were rotated to a common major axis. The best--fit two--dimensional Gaussian model has a FWHM of 0.40$''$$\pm$0.01$''$, consistent with the median FWHM of the individual two--dimensional Gaussian fits. The best--fit S\'ersic profile has a S\'ersic index of $n=1.0\pm0.1$ and an effective radius of $R_e=0.23\pm0.05''$, again indicating that the light profile of the dust emission is consistent with that of an exponential disk.

\begin{figure*}
\includegraphics[angle=180,scale=0.65,trim={0 1.5cm 0 7.2cm},clip]{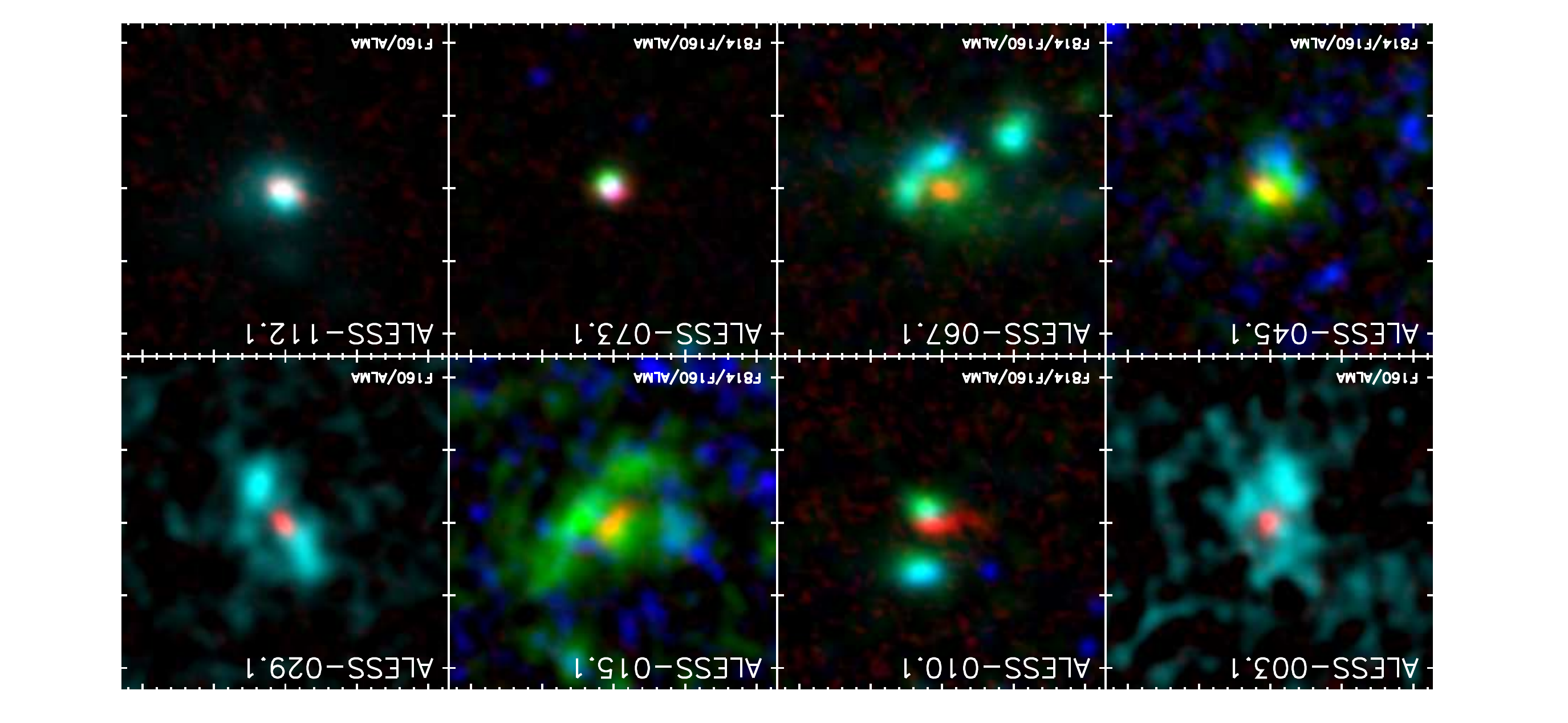}
\caption{False--color images (4.6$''$$\times$4.6$''$ each) constructed from a combination of \textit{HST} and ALMA data for a selection of SMGs with the 
most complete
 data from our sample, showing the 870$\mu$m (red), \textit{I}$_{\rm 814}$--band (green) and \textit{H}$_{\rm 160}$--band imaging (blue). The asymmetric, morphologically complex stellar continuum
 emission appears to be largely uncorrelated with the sites of the significantly more compact and disk--like
 ongoing dusty star formation }
\label{fig:PRthumbs}
\end{figure*}

\subsubsection{$uv$--plane fits}
\label{uvfits}

One way to address whether any flux ``missing'' from the naturally weighted images 
is having an impact on the source sizes measured in the image plane is to measure the sizes directly in the $uv$--plane. Figure~\ref{fig:uvfits} shows the $uv$--data for four isolated ALESS sources. The phase center of the new Cycle~1 datasets have been shifted to center exactly on the ALESS SMGs, and the data have then been radially averaged in bins of 75 k$\lambda$. Also shown are simulated profiles of smooth exponential disks ($n=1$) with the same flux densities, effective radii, and axial ratios as those of the sources, and with added noise. 

To compare these data to the low-resolution Cycle~0 observations, we applied the same procedure to the Cycle~0 data, which have also been scaled by the response of the Cycle~0 primary beam at the position of the SMG. As the majority of the SMGs are unresolved in the Cycle~0 data, only the central data point is shown. 
There is indeed no evidence that the Cycle~1 data are missing any emission, in agreement with \S\ref{flux}. 

We then fit the Cycle~1 $uv$-profiles with two models: 1) a Gaussian, 2) a Gaussian plus a constant. The latter represents a point source (or point sources) in the image plane and was found to be necessary due to the signal evident at large $uv$--distances in the plots (particularly ALESS 17.1). 
We find that this point--source component makes up $\lesssim$5\% of the total emission in ALESS 5.1, 45.1, and 73.1, but it constitutes 15\% of the emission in ALESS 17.1. This is likely caused by the large ellipticity observed in ALESS 17.1, which is nearly unresolved along its minor axis in our map, combined with the fact that the shortest spacings play a larger role in the $uv$--plane fitting. The FWHM values resulting from the Gaussian$+$constant model are listed in Figure~\ref{fig:uvfits}.

These sizes can be most directly compared to one-dimensional (circular) Gaussian fits from azimuthally averaged data in the image plane (Table~\ref{tab:1}). These values tend to be somewhat smaller on average than the 2D elliptical Gaussian fit values (median FWHM$_{1D}$/FWHM$_{2D}$$=$0.79$\pm$0.07), reflecting the ellipticity of the emission observed in the individual sources. When we include a point source component in the $uv$--plane model, we find that the FWHM sizes derived from fitting in the image and $uv$--planes agree (within the uncertainties). From this test and those reported in the \S A\ref{robustness}, we conclude that the sizes measured in the image plane are robust, and that they are unaffected by the presence of any potentially ``missing'' emission.

\begin{figure}
\includegraphics[angle=180,scale=0.37,trim={0 0cm 2cm 2cm},clip]{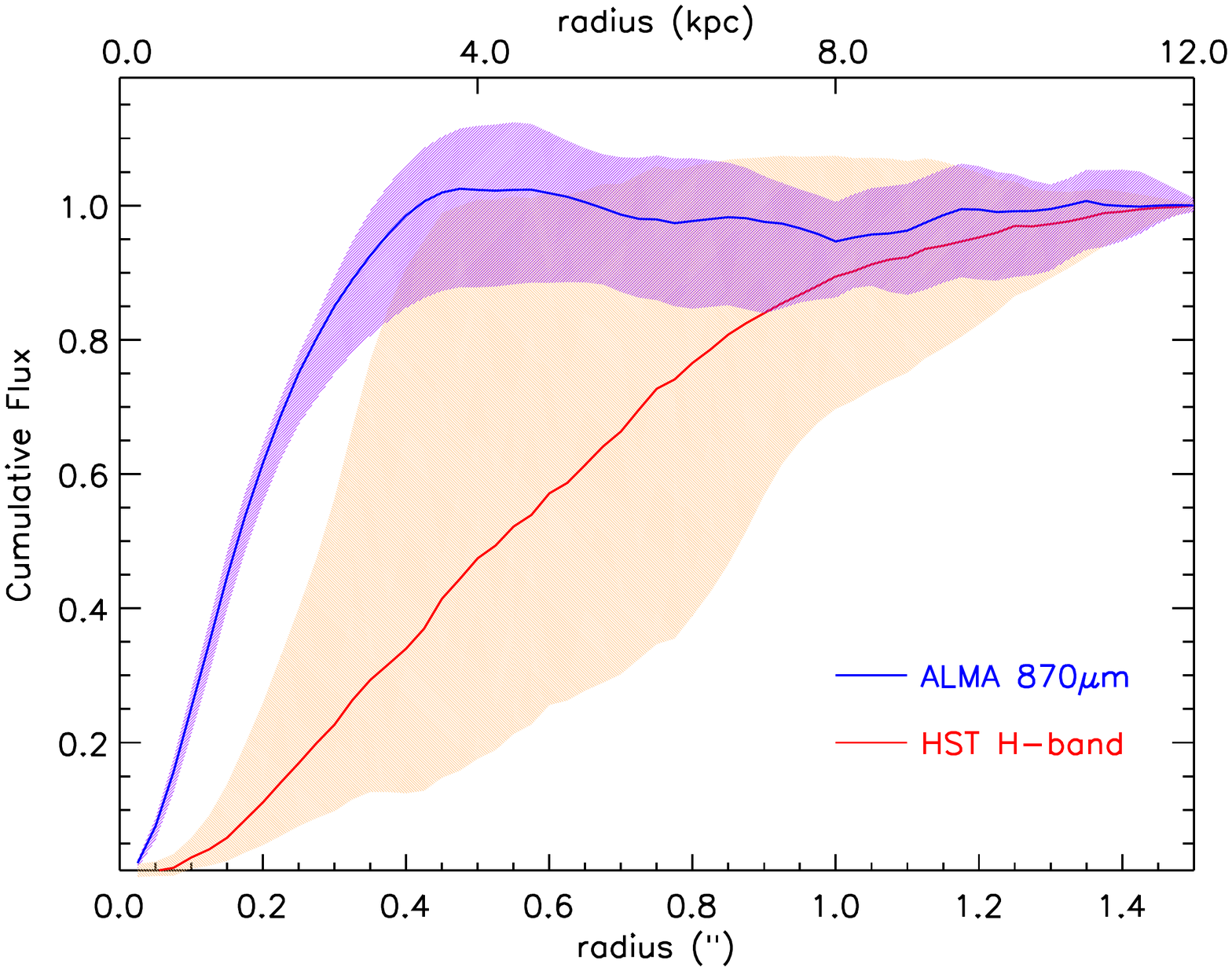}
\caption{Curves of growth for the fraction of the flux density within a 1.5$''$ radius aperture in the ALMA 870$\mu$m and \textit{HST} $H_{\rm 160}$--band images. The solid lines show the median, and the shaded regions show the source--to--source scatter. The top axis denotes the physical scale for a typical redshift of $z\sim2.5$. The obscured star formation traced by the ALMA 870$\mu$m emission appears to be significantly more compact than the unobscured stellar emission. }
\label{fig:cog}
\end{figure}

\begin{figure*}
\centering
\includegraphics[scale=0.8,trim={0.5 18.5cm 0 0},clip]{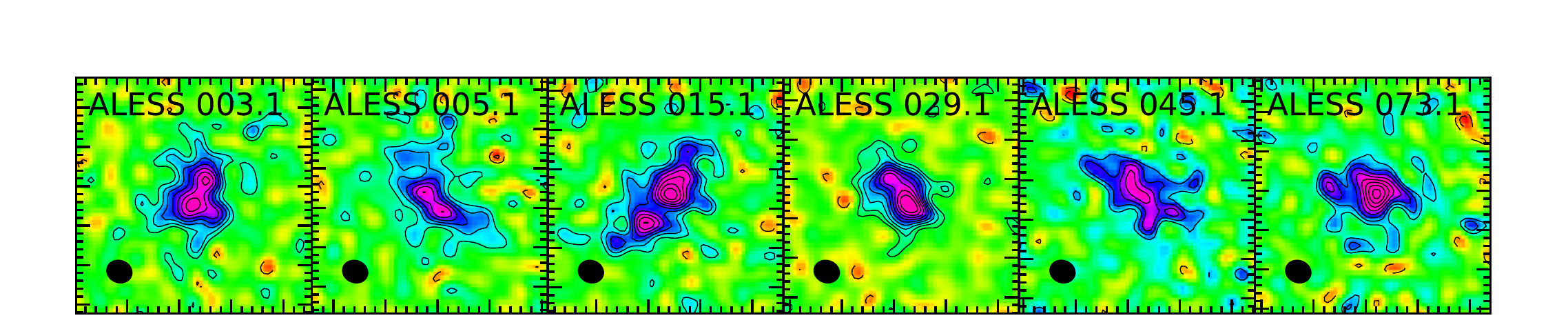}
\vfill
\includegraphics[scale=0.8,trim={0.5 18.5cm 0 1cm},clip]{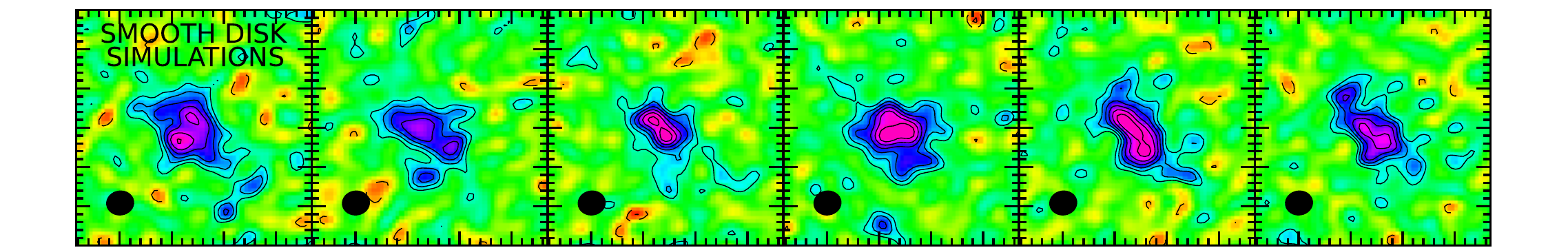}
\caption{\textit{Top row:} Example images (each 1.2$''$$\times$1.2$''$) of the 870$\mu$m emission from our SMG sample imaged with Briggs weighting ($R=-0.5$) to achieve the highest resolution (0.12$''$$\times$0.11$''$). Contours start at $\pm$2$\sigma$ and go in steps of 1$\sigma$, and the typical RMS ($\sigma\sim130$ $\mu$Jy beam$^{-1}$) corresponds to a rest--frame brightness temperature of T$_{B} = 0.4$ K at $z\sim2.5$. \textit{Bottom row:} Simulated observations of {\bf smooth} exponential disks with the same typical angular resolution and S/N as the observations and the same color scaling. This experiment highlights that caution should be exercised when identifying clump--like structure in interferometric maps of even moderate S/N. 
 }
\label{fig:clumps}
\end{figure*}

\subsection{Comparison to stellar emission}
\label{HSTcomp}

Our SMGs were selected to have \textit{HST} WFC3 imaging at comparable (0.15$''$) resolution in one or more bands, providing a less dust--sensitive probe (than optical imaging) of the stellar distribution on $\sim$kpc scales. We tied the astrometry of the \textit{HST} images to the IRAC images, and the relative astrometry between the \textit{HST} and ALMA images is expected to be accurate to $\sim$0.1$''$. False--color images constructed from a combination of the \textit{HST} and ALMA data are shown as multi--band color images for a selection of SMGs with the most complete
 data in Figure~\ref{fig:PRthumbs}, where a variety of stellar morphologies are observed. The 870$\mu$m contours for the full sample are overplotted on the \textit{H}$_{\rm 160}$--band imaging in Figure~\ref{fig:thumbs}. The source positions (and/or stellar environments) of ALESS 5.1 and 10.1 suggest these sources are potentially weakly lensed. In particular, the redshifts of the nearby bright \textit{H}$_{\rm 160}$--band counterparts suggest that these sources are at lower redshift, although we cannot rule out that they are mergers.

It is immediately clear from these comparisons that the obscured star formation traced by the dust emission is generally more compact than the stellar emission. To quantify this effect, the median curves of growth for the naturally weighted ALMA 870$\mu$m maps and \textit{HST} \textit{H}$_{\rm 160}$--band imaging are shown in Figure~\ref{fig:cog}. These growth curves in both cases were calculated using a 1.5$''$ radius aperture centered on the ALMA emission, assuming this is indicating the mass--weighted center of the system. 
 The ALMA 870$\mu$m curve dips below a cumulative fraction of 1.0 at large ($>$0.6$''$) radii due to the presence of negative sidelobes in the ALMA maps. Calculated in this way, the median half--light radius of the ALMA 870$\mu$m emission is 0.16$''$$\pm$0.02$''$, in agreement with the direct integration value given in \S A\ref{robustness}, while the median half--light radius of the \textit{H}$_{\rm 160}$--band imaging is 0.5$''$$\pm$0.1$''$ -- a factor of three larger. 

These comparison also clearly demonstrate the morphological contrast between the internal structure of the obscured and unobscured star formation. While we find evidence that the obscured star formation is distributed in smooth exponential disks at a resolution of $\sim$0.16$''$, the stellar emission on the same scales appears very clumpy and irregular.  \citet{2015ApJ...799..194C} studied the stellar morphologies of a larger sample of 48 ALESS SMGs (including those presented here) and reported that of the $\sim$80\% detected in the \textit{H}$_{\rm 160}$--band down to a median sensitivity of \textit{H}$_{\rm 160}=27.8$ mag, 82$\pm$9\% appear to have disturbed morphologies. This implies that the irregular stellar morphologies we observe are representative of the larger sample.
Based on a statistical comparison with the lower--resolution Cycle~0 data, \citet{2015ApJ...799..194C} also reported an offset between the \textit{H}$_{\rm 160}$--band components and the dusty star--forming regions, which they argued could be due to either obscuration of the rest--frame optical/UV imaging or real misalignment between the dusty star--forming regions and the location of the majority of the unobscured stellar continuum emission
 within the SMGs. They argue that the latter scenario is more likely, given the lack of a difference between the low-- and high--redshift subsamples, as the morphological $K$--correction implies that the rest--frame UV emission traced in higher--redshift sources will be more sensitive to clumpy star--forming regions and dust obscuration. The present comparison demonstrates that the asymmetric, morphologically complex stellar emission indeed appears to be largely uncorrelated with the sites of the ongoing dusty star formation on a case--by--case basis, confirming that the misalignment is real. 

We conclude that the obscured star formation traced by the ALMA 870$\mu$m emission is both significantly smoother and more compact than the unobscured stellar emission. However, it is possible that the resolution of the current ALMA data ($\sim$0.16$''$; $\sim$1.3\,kpc at $z\sim2.5$) is still slightly too coarse to resolve any potential clump--like structure. 
We investigate whether the dust emission shows evidence for clumpy structure as we push down to smaller spatial scales in the next section. 

\begin{figure*}
\centering
\includegraphics[angle=270,scale=0.6,trim={0 0cm 0cm 0cm},clip]{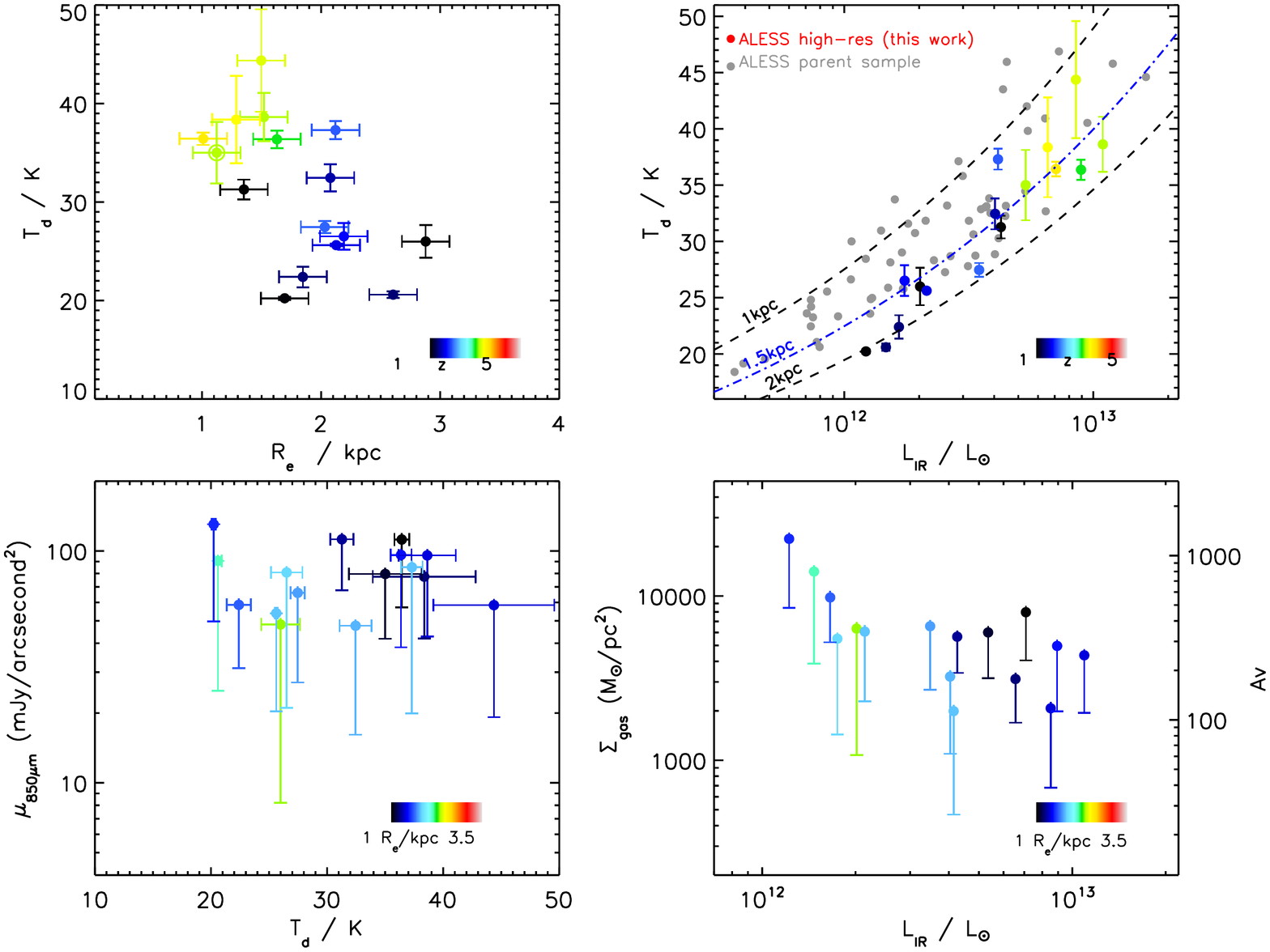}
\caption{\textit{Top left:} The characteristic dust temperature ($T_{\rm d}$) from a modified blackbody fit \citep{2014MNRAS.438.1267S} versus effective radius ($R_{\rm e}$), color coded by source redshift.  The dust emission in the higher--redshift sources appears to be warmer and more compact. 
\textit{Top right:} The characteristic dust temperature versus infrared luminosity ($L_{\rm IR}$) for the targets of this paper compared to the parent ALESS sample. The high--resolution sources (this work) are again color coded by redshift. The dashed lines indicate the physical sizes predicted assuming the Stefan--Boltzmann law. The sizes we derive are consistent with those predicted from this simple model, and there is marginal evidence that the SMGs with the highest luminosities are more compact. 
\textit{Bottom left:} The peak (filled) and average (lower error bar) 870$\mu$m surface brightness from the highest--resolution maps (\S\ref{clumps}) versus characteristic dust temperature, color coded by physical size. There is no trend in surface brightness with dust temperature. 
\textit{Bottom right:} The gas surface density versus infrared luminosity, color coded by physical size. The right--hand axis shows the corresponding visual extinction ($A_{\rm v}$). The gas surface density/extinction values measured are very high, but there appears to be no trend with infrared luminosity. }
\label{fig:Td_FIR}
\end{figure*}

\subsection{Clumps}
\label{clumps}	

Massive ($\sim$10$^{8}$--10$^{10}$\,M$_{\odot}$) kpc--scale star--forming clumps have been argued to be an important feature of high-redshift galaxies, with observational evidence claimed for such clumps in the rest-frame UV \citep[e.g.,][]{2005ApJ...627..632E, 2012ApJ...757..120G}, rest-frame optical \citep[e.g.,][]{2009ApJ...692...12E, 2011ApJ...739...45F}, NIR integral field spectroscopy \citep[e.g.,][]{2008ApJ...687...59G, 2011ApJ...733..101G}, and potentially also CO and rest-frame FIR emission in a handful of the brightest and/or strongly lensed sources \citep[e.g.,][]{2010Natur.463..781T, 2010Natur.464..733S, 2011ApJ...742...11S, 2012ApJ...760...11H, 2016ApJ...827...34O}. 
It has been proposed that these clumps form in--situ from the fragmentation of gravitationally unstable gas disks \citep[e.g.,][]{1998Natur.392..253N, 2009MNRAS.397L..64A, 2012ApJ...757...81B}, though it has also been suggested that some of the most massive clumps may be accreted cores of satellite galaxies \citep[e.g.,][]{2016MNRAS.tmp.1460M, 2016arXiv160303778O},
and reconciling the existence of such clumps with the presence of certain stellar feedback recipes
makes them an important testbed of feedback processes in galaxy formation \citep[e.g.,][]{2016arXiv160606739M}. 
To search for such clumps in our SMGs, we re-image the ALMA 870$\mu$m data with a Briggs robust parameter of  $R=-0.5$, resulting in a resolution of 0.12$''$$\times$0.11$''$. This results in almost a factor of two decrease in beam area over the ``native'' resolution, corresponding to physical scales of 1.0$\times$0.9\,kpc at $z\sim$\,2.5. As a consequence, the typical RMS noise values in the maps approximately doubles to $\sim$130\,$\mu$Jy beam$^{-1}$. 

Figure~\ref{fig:clumps} shows several examples of SMGs imaged in this way, where we have selected those which are clumpiest in appearance. 
It is tempting to conclude from a visual inspection that several of the SMGs break up into a small number of kpc--scale clumps. 
To test this, we used {\sc casa} to simulate 16 observations of smooth exponential disks with the same angular resolution and noise levels as the observations in Figure~\ref{fig:clumps}. The parameters of the input model were tuned to the typical parameters observed in our SMGs: an effective radius of $R_e = 0.26''$, an axis ratio of 0.5, and a total flux density of S$_{870\mu m}\sim6.5$\,mJy.
Several examples of simulated maps are shown in Figure~\ref{fig:clumps} along with the real data, where just as with the real data, we have selected those which are clumpiest in appearance. Indeed, many of the simulated exponential disks break up into a small number of closely-spaced emission peaks, similar to the observed high-resolution maps. This experiment highlights that caution should be exercised when identifying structure in high--resolution interferometric maps at this S/N level (S/N$\sim$5--10).

\begin{figure*}
\includegraphics[angle=0,scale=0.7,trim={1cm 3.3cm 2cm 3.5cm},clip]{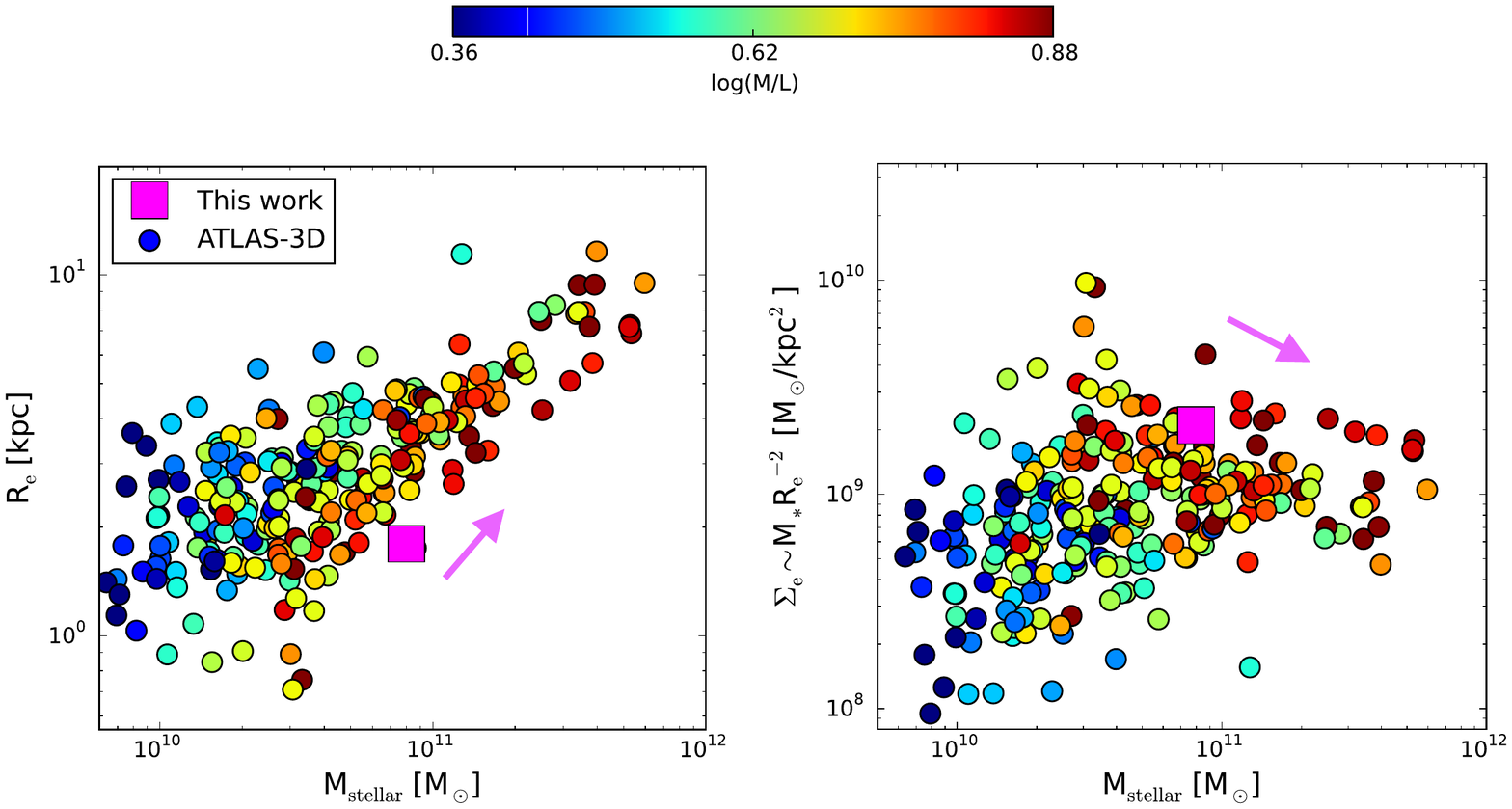}
\caption{Comparison of the ALESS SMG properties to nearby early--type galaxies from the ATLAS$^{\rm 3D}$ project. The ATLAS$^{\rm 3D}$ data (circles color--coded by mass--to--light ratio) come from \citet{2011MNRAS.413..813C, 2013MNRAS.432.1862C}.  The typical properties of the ALESS SMGs are shown, and the arrows indicate the direction that these properties may evolve in with decreasing redshift. The ALESS SMGs studied in this work have stellar masses, effective radii, and average gas surface densities similar to the locus of nearby early--type galaxies, and their descendants are thus expected to have properties similar to the most compact massive early--type galaxies observed locally.}
\label{fig:atlas3d}
\end{figure*}

As a more quantitative analysis, 
we fit each observed SMG with a 2D elliptical Gaussian and subtracted the resulting model of the smooth emission from the high-resolution map. We find that none of the SMGs have residual structure with peak fluxes $>$3$\sigma$. Of the six SMGs with residuals between 2.5--3$\sigma$, the strongest residual (2.9$\sigma$) is due to the possible structure to the East of the main peak in ALESS 73.1 (Figure~\ref{fig:clumps}).

Recognizing that any smooth contribution may be overestimated by this crude method, we note that in all of the sources except for ALESS 15.1, the candidate clumps are only distinct from each other at the 1--2$\sigma$ level even before the subtraction, again consistent with the smooth--disk simulations. The clump candidates in ALESS 15.1 (Figure~\ref{fig:clumps})
 are the only candidates which are separated in brightness by $>$4$\sigma$ in the high-resolution maps. These candidates have peak flux densities of 0.8--1.0\,mJy beam$^{-1}$, integrated flux densities of 2.8--4.2\,mJy, and FWHM areas of 2.3--2.8\,kpc$^2$ \citep[assuming $z_{phot} = 1.93$;][]{2014ApJ...788..125S}. If this structure is real, then scaling the total estimated star formation rate \citep[130\,M$_{\odot}$\,yr$^{-1}$;][]{2014MNRAS.438.1267S} by the ratio of the integrated flux density in each clump over that of the source as a whole gives star formation rate surface densities of $\sim$15--20\,M$_{\odot}$\,yr$^{-1}$\,kpc$^{-2}$ \citep[c.f.,][]{2015ApJ...799...81S}. It is possible that these two clump--like structures are the cores of merging galaxies, though we have no way to distinguish between these scenarios with the current data. We find no strong evidence for corresponding structure in the \textit{HST} \textit{H}$_{\rm 160}$--band image (see Figure~\ref{fig:thumbs}),
  though the counterpart is very faint. 
We conclude that while there may be a hint of clump--like dust emission in the current 870$\mu$m data on $\sim$kpc--scales, higher signal--to--noise observations at higher spatial resolution are required to confirm whether these clumpy structures are indeed real.

\subsection{The $L_{\rm IR}$--$T$ relation and gas surface densities}
\label{correlations}

Infrared--luminous galaxies in the local \citep[e.g.,][]{2009MNRAS.393..653C, 2010MNRAS.409...75H} and high--redshift universe \citep[e.g.,][]{2003MNRAS.338..733B, 2003ApJ...588..186C} have long been known to show a relation between their dust temperatures and infrared luminosities (the $T_{\rm dust}$--$L_{\rm IR}$ relation). This relation is not due simply to selection effects, but is instead a consequence of the Stefan--Boltzmann law relating size, luminosity and dust temperature. We plot this relation as well as the $T_{\rm dust}$--$R_{\rm e}$ relation for our SMGs in Figure~\ref{fig:Td_FIR}, where we have used the $L_{\rm FIR}$ and $T_{\rm dust}$ values reported in \citet{2014MNRAS.438.1267S} for those sources without updated spectroscopic redshifts (Danielson et al. 2016, ApJ, submitted).
 The tracks plotted indicate different physical sizes of a perfect blackbody, and we assume optically thick radiation. We see that the physical scale of the dust emission correlates with redshift and dust temperature. We also see a strong $T_{\rm dust}$--$L_{\rm IR}$ relation implying sizes of 1--2 kpc (with marginal evidence that the dust emission in higher--luminosity SMGs is more compact). The sizes we measure directly from the high--resolution maps (median $R_{e}=1.8\pm$0.2\,kpc) are in agreement with the predictions of this simple model. This result contrasts with the conclusion of \citet{2016ApJ...820L..16Y} based on the modified blackbody equivalent of the Stefan--Boltzmann law applied to strongly lensed sources, where they suggested that the larger sizes measured for their high--redshift sources must be the result of blending. We note that when we use the modified blackbody equivalent of the Stefan--Boltzmann law instead for our sample, the sizes we measure are still consistent with the predictions (median ratio of $R_{\rm eff,predicted}$/$R_{\rm eff,observed}$$=$1.1$\pm$0.2).

We then used the resolved nature of our observations to search for correlations between dust/gas surface density and dust temperature/luminosity. Figure~\ref{fig:Td_FIR} shows 870$\mu$m surface brightness as a function of dust temperature, where we include measurements of both the average and peak surface brightness in each galaxy. No trend is evident between the surface brightness and dust temperature. We have then converted these measurements to gas surface density by scaling the SED--derived dust mass and assuming a gas--to--dust ratio of 100. The corresponding extinction ($A_{\rm v}$) values were calculated as in \citet{2009MNRAS.400.2050G}. The gas surface densities implied by our observations are very high -- over two orders of magnitude higher than GMCs in the nearby universe \citep{1987ApJ...319..730S} -- and similar to those found in local ULIRGs. There appears to be no trend between gas surface density/extinction and total infrared luminosity.

\section{Discussion}
\label{discussion}	
 
The ALMA imaging presented here allows us to resolve the dust-obscured star formation in a sample of luminous
high-redshift dusty star-forming galaxies on scales of $\sim$1\,kpc. S\'ersic profile fits reveal that the galaxies have a median effective radius of $R_e=0.24$$''$$\pm$0.02$''$ at a rest wavelength of $\lambda\sim250\mu$m (for a typical source redshift of $z\sim2.5$), 
 corresponding to a typical physical size of $R_{e}=1.8\pm$0.2\,kpc.
In contrast, \textit{Herschel} 70--160$\mu$m imaging of 400 local galaxies and QSO hosts suggests that ULIRGs are exclusively found with very compact ($R_e\sim0.5$ kpc) morphologies \citep[albeit at shorter rest wavelengths of $\lambda\sim70\mu$m;][]{2016A&A...591A.136L}. This confirms earlier suggestions from CO observations and
marginally resolved radio and submillimeter
data \citep[e.g.,][]{2004ApJ...611..732C, 2010MNRAS.404..198I, 2011MNRAS.412.1913I, 2010ApJ...714.1407C, 2013ApJ...776...22H, 2015ApJ...799...81S, 2015ApJ...810..133I, 2015A&A...584A..32M} 
that high-redshift dusty star-forming galaxies are indeed larger than similarly luminous local galaxies.

In addition to the observed sizes, the observations presented here resolve the dust emission over many beams at relatively high S/N, allowing us to constrain the more detailed morphology. In particular, there have been a number of claims in the literature that, when observed at high--resolution, the gas reservoirs of SMGs break up into sub--kpc or kpc--sized clumps \citep[e.g.,][]{2010Natur.463..781T, 2011ApJ...742...11S, 2012ApJ...760...11H, 2015PASJ...67...93H}. 
Assuming a constant dust--to--gas ratio -- i.e., that the dust follows the gas -- the dust distribution should then be similarly clumpy. Such clumpy dust within a rotating gas disk was potentially observed in, for example, the strongly lensed ``Eyelash'' galaxy by \citet{2010Natur.464..733S}, 
seeming to confirm this theory. 
In contrast, we find that the SMGs observed here appear (within the limits of our current resolution and sensitivity) to be smooth and disk--like on kpc--scales, with a median S\'ersic index of $n=0.9\pm0.2$. 
Combined with the measured sizes ($R_{e}=1.8\pm$0.2\,kpc), this seems to rule out the sort of extended, clumpy disk galaxies predicted by simulations of violent disk instability \citep[e.g.,][]{2009ApJ...703..785D, 2014ApJ...780...57B}
and observed in 
 optically--bright systems \citep[e.g.,][]{2006ApJ...645.1062F}
and potentially even in the ultraluminous $z\sim4$ SMG GN20 \citep{2012ApJ...760...11H, 2015ApJ...798L..18H}. 
The relative uniformity of the dust morphologies observed here also seems to contradict models where SMGs are a heterogenous population \citep[e.g.,][]{2011ApJ...743..159H, 2012MNRAS.424..951H}, although larger sample sizes covering a larger range of flux densities are required to more thoroughly test this conclusion.

It is, of course, still possible that there is clump--like structure below our current resolution limits. The clumps in the Eyelash and SDP.81 are reported to have physical sizes of only a couple hundred pc \citep{2010Natur.464..733S, 2015PASJ...67...93H}. Similarly, the dust continuum in the most well-studied local ULIRG, Arp 220, is concentrated in two very compact ($\sim$30--50\,pc) nuclei situated $\sim$300\,pc apart \citep[although at longer rest-frame wavelengths; e.g.,][]{2008ApJ...684..957S, 2015ApJ...799...10B, 2016arXiv160509381S}. We would not be able to resolve the nuclei of Arp 220 at a redshift of $z\sim2.5$ with the present observations, and indeed, we may find a hint of clump--like structure in one of our SMGs when we push down to (sub--)kpc scales. 
However, the simulations and analysis in \S\ref{clumps} suggest that caution should be exercised when identifying candidate clumps in even moderate S/N interferometric data. 
Indeed, the sizes we measure from the high--resolution images are consistent with those predicted from the Stefan--Boltzmann law based on the the measured dust temperatures and FIR luminosities, another indication that the emission is relatively smooth.
The measured sizes also agree with those estimated from fitting models assuming power--law mass--temperature distributions, again assuming smooth disk emission \citep{2010ApJ...717...29K}.
Significantly
higher--S/N observations at higher resolution are required to determine whether the dust emission in these SMGs retains a disk--like appearance on sub--kpc scales. 

In contrast to the smooth appearance of the obscured star formation, the matched--resolution \textit{HST} WFC3 imaging of these SMGs -- tracing the unobscured rest--frame optical light -- appears clumpy and irregular. The median half--light radius observed for the unobscured stellar emission in these sources corresponds to $R_e=4.1\pm0.8$\,kpc at $z\sim2.5$, implying that the pre--existing
stellar distributions of the SMGs are also significantly more extended than the dust emission. A similar conclusion was drawn regarding the morphology and extent of the stellar component for the larger sample of 48 ALESS SMGs presented by \citet{2015ApJ...799..194C}, indicating that stellar morphologies observed in our sources are representative of the parent population. The current study reveals that this unobscured  stellar emission
 is largely uncorrelated with the obscured star-forming regions in individual sources. This observation implies that SED fitting routines assuming a simple dust screen over a single or even
 composite stellar population may be too simplistic. 

The difference observed between the morphology of the obscured star formation 
and unobscured stellar emission
 in these SMGs also leads us to consider their formation scenario. 
\citet{2015ApJ...799..194C} use the apparently disturbed rest--frame optical morphologies, along with the short expected lifetimes of SMGs, to argue that the majority of $z\sim2-3$ SMGs are early/mid--stage major mergers, as has been argued previously on the basis of, e.g., radio and submillimeter multiplicity and kinematics \citep[e.g.,][]{2006MNRAS.371..465S, 2010ApJ...724..233E}.
 Theoretically, the profiles of merger remnants are expected to be relatively compact and strongly centrally peaked due to the violent and dissipative collapse expected in turbulent and clumpy gas \citep[e.g.,][]{2011ApJ...730....4B}. The small sizes of the dust disks we measure could be consistent with this scenario, though the observed S\'ersic indices are lower than expected in the simulations.

If the starbursts in these galaxies are major merger driven, we are likely observing the result of the gas/dust more rapidly (re--)forming disk structures than the existing stellar component. Assuming a typical gas consumption timescale for SMGs of $\sim$100 Myr \citep{2013MNRAS.429.3047B}, and based on the apparent dynamical (orbital) timescales ($\sim$20 Myr) implied assuming velocity widths of $\sim$500 km s$^{-1}$ \citep{2013MNRAS.429.3047B} and the effective radii measured here, it is possible that the disks have settled while the burst of star formation is still ongoing. 
It is possible that the more compact stellar counterparts observed in some sources (Figure~\ref{fig:PRthumbs}) then correspond to more evolved systems. 
Simulations show that the old stars present in the existing stellar component may also contract due to the turbulent dissipation of the gas and young stars, which can contain a large fraction of the total mass \citep{2011ApJ...730....4B}. 
The current bursts of star formation thus have the potential to transform both the observed galaxy sizes and the overall light profiles as they evolve.

This transformation could also help establish the connection between SMGs and local elliptical galaxies, their proposed descendants \citep[e.g.,][]{1999ApJ...515..518E, 2006MNRAS.371..465S, 2015ApJ...810..133I}. 
In Figure~\ref{fig:atlas3d}, we compare the properties of the ALESS SMGs studied in this work with the volume--limited ATLAS$^{\rm 3D}$ sample of nearby early--type galaxies \citep{2011MNRAS.413..813C}. The stellar masses, effective radii, and mass surface densities for the ATLAS$^{\rm 3D}$ galaxies are discussed in \citet{2013MNRAS.432.1862C}. The median properties\footnote{We show the median properties of the ALESS SMGs as there can be significant scatter among individual galaxies.} of the ALESS SMGs from this work are overplotted, where we use the average gas mass surface densities (Figure~\ref{fig:Td_FIR}). 
 If we assume an average stellar mass of M$_{*}\sim$ 8$\times$10$^{10}$ M$_{\odot}$ \citep{2014ApJ...788..125S}
and a gas mass of M$_{\rm gas}\sim$ 5$\times$10$^{10}$ M$_{\odot}$ (\citealt{2013MNRAS.429.3047B}; consistent with that derived from the dust masses for our sources), 
then the $z\sim0$ descendants of these SMGs would have total masses of M$_{*}\sim$ 1--2$\times$10$^{11}$ M$_{\odot}$ (assuming $\sim$100\% star formation efficiency in the disk). 
If we then assume $z\sim0$ sizes of $R_{\rm e} \sim$ 2--3 kpc \citep[taking the weighted average of the submillimeter and optical sizes, and assuming the stellar components may also contract further;][]{2011ApJ...730....4B}, we can estimate how the descendants of SMGs may compare to local early--type galaxies. We find that the SMG descendants have stellar masses, effective radii, and average gas surface densities consistent with the most compact massive (M$_{*}\sim$ 1--2$\times$10$^{11}$ M$_{\odot}$) early--type galaxies -- with the highest M/L ratios --
 observed locally (Figure~\ref{fig:atlas3d}).

\section{Summary}
\label{summary}	

We have presented high-resolution ($\sim$0.16$''$; $\sim$1.3\,kpc at $z\sim2.5$) 870$\mu$m ALMA imaging of 16 luminous
 ALESS SMGs, allowing us to clearly resolve the dust-obscured star formation in these $z\sim2.5$ galaxies on $\sim$1 kpc scales. 
The median light profile has an effective radius of $R_e=0.24$$''$$\pm$0.02$''$ (corresponding to a typical physical size of $R_{e}=1.8\pm$0.2\,kpc) and a S\'ersic index of $n=0.9$$\pm$0.2, implying that the dust emission and, by implication, the
 obscured star formation is remarkably disk-like at the current resolution and sensitivity.
We present a series of tests in the image and $uv$--planes to confirm that the fraction ($\sim$10--15\%) of emission that may be potentially ``missing'' from the naturally weighted maps does not bias our conclusions regarding the light profiles or sizes. Our results confirm earlier suggestions that high-redshift dusty star-forming galaxies are indeed larger than similarly luminous local galaxies.

We find that the present observations paint a different picture to the disturbed morphologies observed in the stellar distributions of the SMGs traced by \textit{HST} \textit{H}$_{\rm 160}$-band imaging. In particular, the extended, morphologically complex stellar emission appears to be largely uncorrelated with the sites of the ongoing dusty star formation. This observation has implications for SED fitting routines assuming a simple dust screen over a single composite stellar population. 

To search for clump--like structure in the dust--obscured star formation, we use different weighting schemes with the visibilities to probe scales of 0.12$''$ (1.0\,kpc), but we find no significant evidence for clumping in the majority of cases.  Indeed, we demonstrate that the observed morphologies are generally consistent with those seen in
simulated interferometric images of smooth exponential disks at similar (moderate) S/N. 
 This experiment highlights that caution should be exercised when identifying structure in high--resolution interferometric maps at this S/N level (S/N$\sim$5--10). While the present observations suggest that kpc--scale clumps of dust (and cool gas) are rare in these systems, higher--S/N observations of the dust-obscured star formation and molecular gas at higher resolution will be crucial in order to test whether the apparently smooth dust (and by implication, gas) distribution becomes more structured on sub--kpc scales. 

We examine a number of correlations between physical parameters for these SMGs, including the well-known $T_{\rm dust}$--$L_{\rm IR}$ relation, and we find that the source sizes we measure directly from the high--resolution maps are consistent with those predicted by this simple relation. This agreement is another indication that the emission is relatively smooth. While the physical scale of the dust emission appears to correlate with dust temperature and redshift, no trend is evident between the surface brightness and dust temperature, nor between gas surface density/extinction and total infrared luminosity. The gas surface densities implied by our observations are significantly higher than GMCs in the nearby universe, and similar to those found in local ULIRGs. 

The lack of clumps in the obscured star formation, in combination with the compact sizes, seems 
to rule out the sort of extended, clumpy disk galaxies predicted by simulations of violent disk instability \citep[e.g.,][]{2009ApJ...703..785D, 2014ApJ...780...57B}. The compact nature of the obscured star formation compared to the existing stellar component may instead suggest that the bursts are fueled by major mergers, although the exponential light profiles we observe are seemingly inconsistent with the spheroids that are thought to result from the highly dissipative collapse. The relative uniformity in the observed dust morphologies may contradict suggestions of a heterogeneous SMG population, although larger samples of galaxies covering a wider
 range of flux densities are required to thoroughly test this conclusion, as these models suggest that the observed morphology is a function of SMG flux density.

Given the stark contrast between the observed dust and stellar morphologies, we suggest that the current bursts of star formation have the potential to transform both the observed galaxy sizes and the overall light profiles as they evolve.
This transformation could help establish the connection between high--redshift SMGs and red--and--dead local elliptical galaxies, their proposed descendants. 
We compare the observed properties of our SMGs to the volume--limited sample of ATLAS$^{\rm 3D}$ nearby early--type galaxies, and we suggest that the likely $z\sim0$ descendants of SMGs have average properties -- including stellar masses, effective radii, and gas surface densities -- that are consistent with the most compact massive (M$_{*}\sim$ 1--2$\times$10$^{11}$ M$_{\odot}$) early--type galaxies observed locally.

\acknowledgements
The authors wish to thank both the NAASC and the Allegro ARC node, including Todd Hunter, Alison Peck, and Michiel Hogerheijde, for assistance with collecting and analyzing the ALMA data.
We also thank Romeel Dav\'e, Marijn Franx, and Nick Scoville for useful discussion and the referee for the positive report. This work was performed in part at the Aspen Center for Physics, which is supported by National Science Foundation grant PHY-1066293.
IRS acknowledges support from STFC (ST/L00075/1), the ERC Advanced Grant {\sc dustygal} (321334) and a Royal Society/Wolfsom Merit Award. 
RJI acknowledges support from the European Research Council in the form of the Advanced Investigator Program, 321302, {\sc cosmicism}.
FW acknowledges ERC Starting Grant Cosmic$\_$Dawn. This paper makes use of the following ALMA data:  ADS/JAO.ALMA\#2011.1.00294.S and ADS/JAO.ALMA\#2012.1.00307.S. 
AK acknowledges support by the Collaborative Research Council 956, sub--project A1, funded by the Deutsche Forschungsgemeinschaft (DFG).
KK acknowledges support from the Swedish Research Council. 
This work was performed at the Aspen Center for Physics, which is supported by National Science Foundation grant PHY-1066293. ALMA is a partnership of ESO (representing its member states), NSF (USA) and NINS (Japan), together with NRC (Canada) and NSC and ASIAA (Taiwan) and KASI (Republic of Korea), in cooperation with the Republic of Chile. The Joint ALMA Observatory is operated by ESO, AUI/NRAO and NAOJ.
The National Radio Astronomy Observatory is a facility of the National Science Foundation operated under cooperative agreement by Associated Universities, Inc.


\appendix
\renewcommand\thefigure{\thesection\arabic{figure}}    

\subsection{A.1. Robustness of parameters}
\label{robustness}

While the analysis in \S\ref{flux} shows that the low-resolution Cycle~0 estimated flux density for each SMG is recovered in the new data, the difference between the $uv$--tapered and naturally weighted (0.16$''$ FWHM synthesized beam) images of the new data implies that the latter may be insensitive to a fraction ($\sim$10--15\%) of the emission from what is presumably a more extended component. In order to test whether this is affecting the parameters derived in the previous section, we have carried out a series of tests in the image plane. 

As the first test, we fit two-dimensional Gaussian and S\'ersic profiles to the data $uv$--tapered to 0.3$''$. This tapering should recover the flux potentially ``missing'' from the naturally weighted maps (see \S\ref{flux}) but present on the shortest baselines
without degrading the image quality more than necessary. The Gaussian fits have a median major axis size of FWHM$=$0.42$''$$\pm$0.04$''$, and the S\'ersic profile fits have a median index of $n=0.9$$\pm$0.3 and an effective radius of $R_e=0.21$$''$$\pm$0.05$''$. These values are all consistent with the profiles derived from the naturally weighted maps.

As a second test, we computed the half--light radii for the sources in the naturally weighted maps by simply determining the radius within which half the light is contained -- i.e., with no preference for a particular profile. We then repeated this exercise using the total flux estimates from the $uv$--tapered maps in the denominator, and conservatively assuming that this flux lies entirely outside the measured radii. The median half--light radii determined in this manner are $R_e=0.18$$''$$\pm$0.02$''$ and $R_e=0.19$$''$$\pm$0.02$''$, respectively -- showing excellent agreement. 

As a third test, we took the naturally weighted images and added 15\% of the emission in a 1$''$--diameter uniform disk around each source. We then re-fit the images with two--dimensional Gaussian and S\'ersic profiles. The resulting Gaussian fits have a median major axis size of FWHM$=$0.45$''$$\pm$0.03$''$, and the S\'ersic profile fits have a median index of $n=1.0$$\pm$0.3 and an effective radius of $R_e=0.26$$''$$\pm$0.01$''$. These results are again consistent with the values measured from the naturally weighted maps, indicating no significant bias in the measured properties as a result of any flux potentially ``missing" from the images up to the maximum estimated fraction of 10--15\%.

\renewcommand{\thefigure}{A\arabic{figure}}
\setcounter{figure}{0}    

\begin{figure*}
\begin{center}
\includegraphics[trim={0cm 6cm 1cm 2cm},clip,scale=0.9]{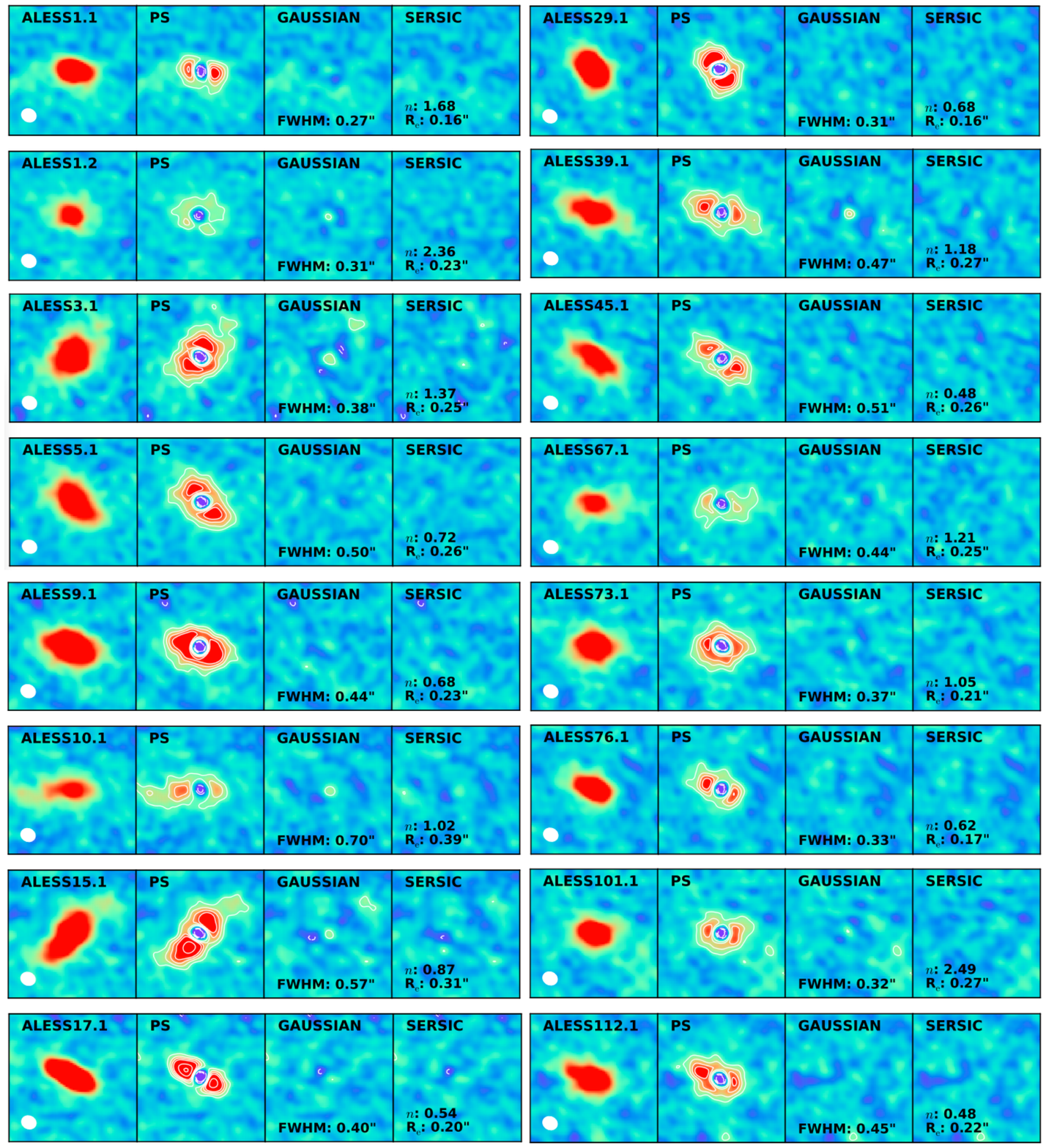}
\caption{Images (1.5$''$$\times$1.5$''$ each) showing the image--plane fitting of the emission profiles observed in the high--resolution ALMA data, including the naturally weighted images (left panel; 0.17$''$$\times$0.15$''$ resolution) as well as the residuals from fitting each source with a point source (PS), two--dimensional Gaussian, and S\'ersic profile. Contours in the residual panels indicate $\pm$3,5,7...$\sigma$. The point source fit is ruled out by $>$5$\sigma$ residuals in all cases, and we see no strong preference between Gaussian and S\'ersic models.}
\vspace{2mm}
\label{fig:modelfits}
\end{center}
\end{figure*}

\begin{figure*}
\includegraphics[angle=180,scale=0.7,trim={1cm 1cm 1cm 7cm},clip]{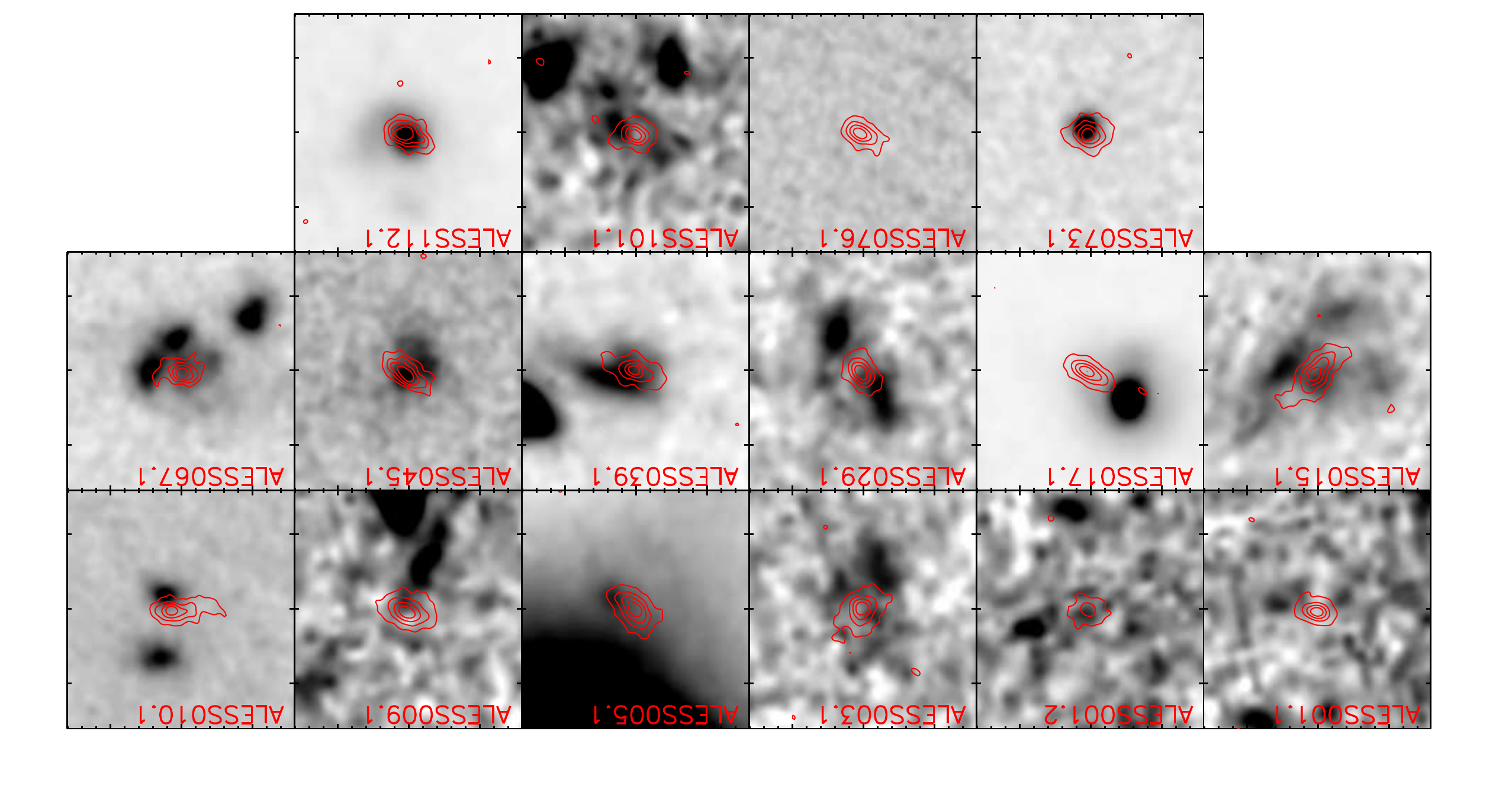}
\caption{Comparisons (each 3.2$''$$\times$3.2$''$) of the ALMA 870$\mu$m emission (contours) and \textit{HST} $H_{\rm 160}$--band emission (grayscale) at similar ($\sim$0.15$''$) resolution for our SMGs. As ALESS 76.1 is not covered by the $H_{\rm 160}$--band imaging, the grayscale for this source shows \textit{I}$_{\rm 814}$--band imaging instead. A variety of stellar morphologies are observed, and the stellar environments suggest that ALESS 5.1 and 10.1 are potentially weakly lensed by nearby foreground bright galaxies (see \S\ref{HSTcomp}). }
\label{fig:thumbs}
\end{figure*}


\bibliographystyle{apj}		
\bibliography{ALESScyc1_SMGstructure_ArXiv}



\end{document}